\documentclass{article}
\usepackage[T1]{fontenc}
\usepackage[latin9]{inputenc}
\usepackage[letterpaper]{geometry}
\usepackage{textcomp}
\usepackage{amstext}
\usepackage{graphicx}
\usepackage{esint}
\usepackage[authoryear]{natbib}

\makeatletter
\newcommand{\lyxaddress}[1]{
	\par {\raggedright #1
	\vspace{1.4em}
	\noindent\par}
}

\makeatother

\begin{document}
\title{Combined magnetic and gravity measurements probe the deep zonal flows of the gas giants}
\author{Eli Galanti and Yohai Kaspi,
\\
(Monthly Notices of the Royal Astronomical Society, accepted)}
\maketitle

\lyxaddress{\begin{center}
\textit{Department of Earth and Planetary Sciences, Weizmann
Institute of Science, Rehovot, Israel}
\par\end{center}}
\begin{abstract}
\textbf{During the past few years, both the Cassini mission at Saturn and
the Juno mission at Jupiter, provided measurements with unprecedented
accuracy of the gravity and magnetic fields of the two gas giants.
Using the gravity measurements, it was found that the strong zonal
flows observed at the cloud-level of the gas giants are likely to
extend thousands of kilometers deep into the planetary interior. However,
the gravity measurements alone, which are by definition an integrative
measure of mass, cannot constrain with high certainty the exact vertical
structure of the flow. Taking into account the recent Cassini magnetic
field measurements of Saturn, and past secular variations of Jupiter's
magnetic field, we obtain an additional physical constraint on the
vertical decay profile of the observed zonal flows on these planets.
Our combined gravity-magnetic analysis reveals that the cloud-level
winds on Saturn (Jupiter) extend with very little decay, i.e., barotropically,
down to a depth of around 7,000~km (2,000~km) and then decay rapidly
in the semiconducting region, so that within the next 1,000 km (600~km)
their value reduces to about 1\% of that at the cloud-level. These
results indicate that there is no significant mechanism acting to
decay the flow in the outer neutral region, and that the interaction
with the magnetic field in the semiconducting region might play a
central role in the decay of the flows.}
\end{abstract}

\section{Introduction}

\begin{figure}
\begin{centering}
\includegraphics{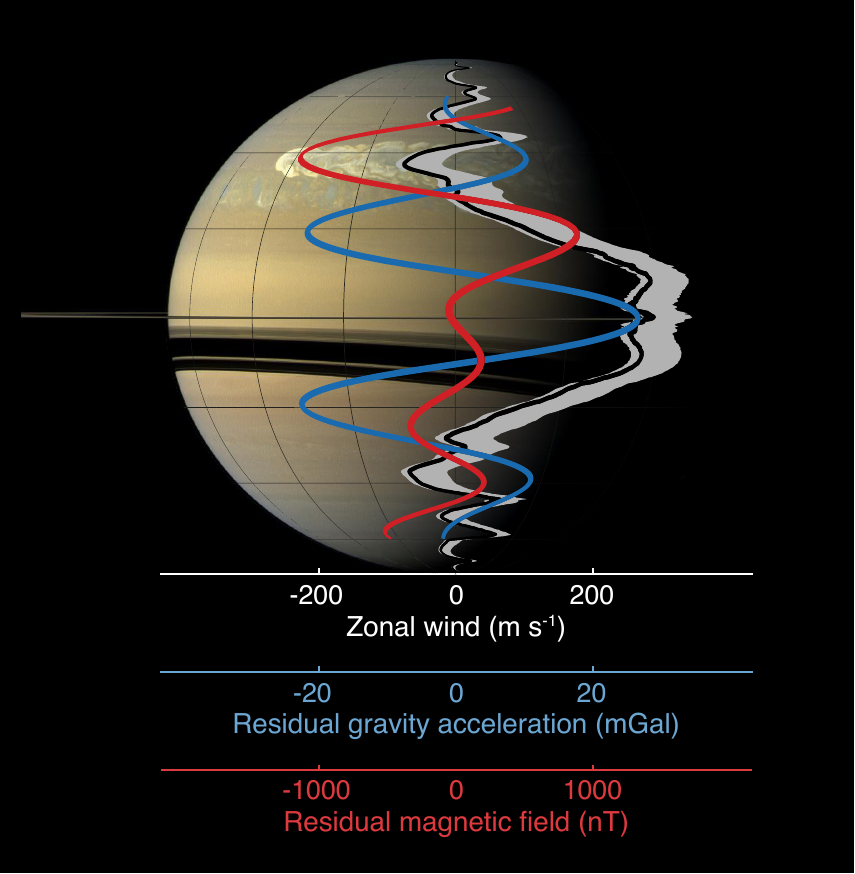}
\par\end{centering}
\centering{}\caption{\textbf{The three measurements used in this study to calculate the
Saturnian flow structure.} The cloud-level wind \citep{Garcia-Melendo2011}
adjusted with the recently estimated rotation period of 10h~34m \citep{Helled2015,Mankovich2019}
(black) and a range of winds calculated for rotation periods between
10h~32m \citep{Read2009} and 10h~39m \citep{Smith1982} (white
shading). The residual gravity field at the planet's surface $R_{S}$
(blue) based on the dynamical contribution to gravity harmonics $J_{3}$
and $J_{5}$ to $J_{10}$ \citep{Iess2019,Galanti2019a}. The residual
radial magnetic field at $0.875R_{S}$ (red) based on the measured
Gauss coefficients $g_{4}^{0}$ to $g_{11}^{0}$ \citep{Dougherty2018,Cao2020}.
All values composing the gravity and magnetic fields appear in Table~\ref{tab:table1}.
Measurements are shown on top of a Saturn picture taken during the
Cassini mission (NASA/JPL-Caltech).\label{fig:Measurements}}
\end{figure}

The strong east-west zonal winds at the cloud-level of Jupiter and
Saturn have been observed to be largely stable over the past several
decades, based on the detection of cloud motion \citep{Sanchez-Lavega2000,Porco2003,Garcia-Melendo2011,Tollefson2017}.
The winds on Jupiter are organized in alternating zonal jets that
reach ${\rm \sim140\,m\,s^{-1}}$ at low latitudes, with a strong
asymmetry between the jets around latitude 20$^{\circ}$ North and
South. The winds on Saturn are mostly hemispherically symmetric, with
a wide, strong eastward flow of nearly ${\rm 300\,m\,s^{-1}}$ at
the equatorial region, and alternating midlatitude jets that are weaker
and less hemispherically symmetric (Figure~\ref{fig:Measurements},
black). On both planets, the observations carry uncertainties from
different sources, of up to $\text{\ensuremath{\pm}}50\,{\rm m\,s^{-1}}$
on Saturn \citep{Garcia-Melendo2011}, and up to $\text{\ensuremath{\pm}}20\,{\rm m\,s^{-1}}$
on Jupiter \citep{Tollefson2017,Fletcher2020}. In addition, since
the winds are measured relative to some reference rotation rate, uncertainty
in Saturn's spin implies a possible range of the wind velocities,
depending on whether the Voyager-based rotation \citep{Smith1982}
or the more recently calculated faster rotation rates \citep{Helled2015,Mankovich2019}
are used (Figure~\ref{fig:Measurements}, white shading). In spite
of the multitude of observations of the cloud-level winds, the only
direct measurement below the cloud-level comes from the Galileo probe
at Jupiter, which showed that at latitude 6.5$^{\circ}$N the zonal
wind increases with depth in the top few bars, and then remains nearly
constant (barotropic) down to 21 bars (130~km deep)\citep{Atkinson1996}.

Recently, the latitudinally-dependent gravity fields of Jupiter and
Saturn were measured with high accuracy by the Juno and Cassini spacecraft,
respectively. On Jupiter, the gravity field was found to have significant
hemispherical asymmetries \citep{Iess2018}, which can be explained
by the cloud-level winds extending deep into the planet \citep{Kaspi2013a,Kaspi2018}.
On Saturn, even the symmetric part of the gravity field was found
to differ substantially from that predicted with a rotating rigid-body,
especially for the higher gravity harmonics \citep{Iess2019} (Figure~\ref{fig:Measurements},
blue). This difference was attributed to the winds extending thousands
of kilometers deep \citep{Galanti2019a}. For both planets, the gravity
measurements indicate not only the overall depth of the winds but
also that the same meridional profile of the zonal flows likely extends
to these depths \citep{Kaspi2020}.

The measurements of the gravity field, which is by definition an integrative
measure of mass, cannot constrain with high certainty the detailed
vertical structure of the flow, since different distributions of density
anomalies can be expressed in the same gravity field at the planet's
surface. Therefore, in both planets, the solutions for the flow fields
discussed above are not unique, and other solutions that are not tied
to the cloud-level winds, can be found to give an exact match to the
gravity measurements, as discussed for Jupiter \citep{Kong2018} and
Saturn \citep{Qin2020}. However, given the small number of gravity
measurements in both planets, and the fact that taking the observed
cloud-level flows and extending them into the interior in a simple
way matches both in sign and in magnitude the measured gravity harmonics
gives a good indication that indeed the interior flows resemble those
at the cloud level. Furthermore, the solutions suggested in these
studies exhibit a flow of the order of $1\,{\rm m\,s^{-1}}$ at depths
of $0.2R$ for Jupiter and $0.3R$ for Saturn. However, while not
fully determined, current estimates of the conductivity in these regions
\citep{Liu2008,Wicht2019a} would imply a generation of a very strong
Lorentz force, that would act to diminish the flow \citep{Cao2017a,Moore2019,Duer2019}.
Finally, in a recent study, a wide range of possible flow structures
was examined in the context of Jupiter gravity measurements \citep{Duer2020}.
Varying both the cloud-level wind profiles and the way the wind decays
with depth, it was found that while the measured gravity field can
be explained by various flow structures, within the range examined
in the study, solutions that differ considerably from the observed
cloud-level winds are statistically unlikely.

Both missions have also measured at an unprecedented accuracy the
magnetic fields of both planets, which differ dramatically between
the two planets. On Jupiter, the field has a complex latitudinal and
longitudinal structure \citep{Connerney2018}, a characteristic that
might be exploited to constrain the flow structure using magnetic
secular variations \citep{Moore2019,Duer2019}. Conversely, the magnetic
field on Saturn is extremely axisymmetric \citep{Dougherty2018,Cao2020},
with latitudinal variability reflecting not only low-order harmonics
but also the contribution from higher harmonics (Figure~\ref{fig:Measurements},
red), which may be related to the structure of the flow below the
cloud-level \citep{Gastine2014,Cao2017a}. 

On both planets, the magnetic field measurements provide valuable
information that can potentially be used to further constrain the
structure of the zonal winds below the cloud-level. Here we report,
for the first time, on the well-confined structure of the flow field
of Saturn, calculated based on the measured cloud-level wind and \textit{both}
the gravity and magnetic measurements. We then extend our analysis
to include the structure of Jupiter's flow field, using both the measured
gravity field and the estimated secular variations of the measured
magnetic field.

\section{Methods}

\subsection{The thermal wind balance\label{subsec:TW}}

Large-scale flows on rapidly rotating planets have a direct relation
to density anomalies and, therefore, affect the gravity field if the
flows are deep enough (i.e., involve a large mass) \citep{Hubbard1999,Kaspi2010a}.
Such a flow is governed by a geostrophic balance between the anomalous
pressure gradient and the Coriolis force \citep{Pedlosky1987,Kaspi2009}.
Given that on Saturn and Jupiter the flow is predominantly zonally
symmetric and assuming sphericity \citep{Galanti2017a}, the resulting
vorticity dynamical balance is between the flow gradient in the direction
parallel to the axis of rotation and the meridional gradient of density
perturbations, known at thermal wind (TW) balance, given by

\begin{equation}
2\Omega r\frac{\partial}{\partial z}\left(\rho_{0}u\right)=g_{0}\frac{\partial\rho'}{\partial\theta},\label{eq: thermal wind}
\end{equation}
where $u(r,\theta)$ is the zonal zonal flow field, $\Omega$ is the
planet's rotation rate, $\rho_{0}(r)$ and $g_{0}(r)$ are the rigid-body
density and gravity fields, $\rho'(r,\theta)$ is the anomalous density
field, and $z$ is the direction of the axis of rotation. Note that
this not the standard atmospheric form of the thermal wind equation
\citep{Holton1992} as the derivative on the left-hand-side is not
in the radial direction, but in the direction of the spin axis \citep{Kaspi2009}.
Other effects not included in this balance, such as the anomalous
gravity and centrifugal forces induced by the density anomalies \citep{Zhang2015,Cao2017b},
were shown, for the large scale zonal flows, to have a small effect
on the gravity solutions \citep{Galanti2017a,Kaspi2018}, and therefore
are not taken into account here. For the background density $\rho_{0}(r)$
we use the same profiles as were used in the gravity-only studies
for Saturn \citep{Galanti2019a} and Jupiter \citep{Kaspi2018}.

The anomalous density field $\rho'$ can then be used to calculate
the wind-induced gravity harmonics
\begin{equation}
\Delta J_{n}^{{\rm mod}}=-\frac{2\pi}{MR^{n}}\intop_{-\pi/2}^{\pi/2}\cos\theta d\theta\intop_{0}^{R}r{}^{n+2}drP_{n}\left(\sin\theta\right)\rho'\left(r,\theta\right)dr,\label{eq: Jn model-1}
\end{equation}
where $\Delta J_{n}^{{\rm mod}},\,n=2,...,N$ are the coefficients
of the wind induced gravity harmonics, $R$ is the planetary radius,
and $M$ is the planetary mass. This principle was successfully used
to calculate the overall depth of the winds on Jupiter and Saturn
using the Juno and Cassini measured gravity field \citep{Kaspi2018,Iess2019,Galanti2019a}.
For the case of Jupiter the calculation was based on the odd gravity
harmonics only \citep{Kaspi2018}, and for the case of Saturn also
the even harmonics were used, with the expected rigid-body solution
subtracted from the measurements (Figure~\ref{fig:Measurements},
blue, see also Table~\ref{tab:table1}a). These are the solutions
we reinvestigate in this study with the new constraints from the magnetic
field measurements.

\subsection{The mean-field electrodynamic balance\label{subsec:MFED}}

The large scale flow field in Saturn and Jupiter can be also related
to a residual magnetic field, induced by the flow in the semi-conducting
region where the fluid begins to become conductive \citep{Galanti2017e,Cao2017a,Duer2019}.
The steady-state balance between the residual magnetic field and the
flow, named the mean-field electrodynamics (MFED) balance \citep{Cao2017a},
is
\begin{eqnarray}
\eta_{E}\left(\nabla^{2}-\frac{1}{s^{2}}\right)B+\frac{1}{r}\frac{d\eta_{E}}{dr}\frac{\partial(rB)}{\partial r} & = & -{\bf B_{0}}\cdot{\bf \nabla}u,\\
\eta_{E}\left(\nabla^{2}-\frac{1}{s^{2}}\right)A & = & -\alpha B,\label{eq:MFED-equations}
\end{eqnarray}
where $A(r,\theta,t)$ and $B(r,\theta,t)$ compose the residual magnetic
field $\mathbf{B}=\nabla\times(A\hat{e}_{\phi})+B\hat{e}_{\phi}$,
$\mathbf{B_{0}}=B_{0}^{r}\hat{e}_{r}+B_{0}^{\theta}\hat{e}_{\theta}$
is the background planetary magnetic field, $\eta_{E}(r)$ is the
effective magnetic diffusivity which is inversely proportional to
the electrical conductivity $\sigma$, $s=r\sin\theta$ is the distance
from the axis of rotation. The function $\alpha(r,\theta)=\alpha_{0}\frac{\eta_{0}}{\eta}{\rm erf}\left(\frac{\theta}{0.005\pi}\right)$
is the dynamo $\alpha$-effect \citep{Cao2017a}, where $\alpha_{0}=10^{-4}\,{\rm m\,s^{-1}}$
is the value at the base of the semi-conducting region (sets by taking
the convective velocity there as 1~mm~s$^{-1}$, and assuming the
effective dynamo alpha-effect is 10\% of the velocity), and $\eta_{0}$
is the value of the magnetic diffusivity at the base of the semiconducting
region. 

In this study, we set the electrical conductivity $\sigma$ as an
analytical function \citep{Cao2017a} that reproduces well the measured
values \citep{Liu2008,French2012}. The outer boundary is set where
$\sigma=10^{-4}$~S~m$^{-1}$ ($0.9R_{S}$ for Saturn and $0.98R_{J}$
for Jupiter), and the inner boundary is set where $\sigma=10^{3}$~S~m$^{-1}$
($0.845R_{S}$ for Saturn and $0.93R_{J}$ for Jupiter) following
\citet{Cao2017a}. The transition depth is set where $\sigma=0.01$~S~m$^{-1}$,
resulting in $R_{T}=0.875R_{S}$ for Saturn and $R_{T}=0.972R_{J}$
for Jupiter. Note that the transition depth $R_{T}$ might be defined
differently \citep{Wicht2019a}, however, this should not affect substantially
our results, as long as the conductivity profile remains the same.
The model solution is given in terms of the Gauss coefficients $g_{n}^{0}$,
similar to the measurements (Table~\ref{tab:table2}). The relation
between the Gauss coefficients and the latitude dependent magnetic
field in the radial direction \citep{Dougherty2018}, estimated at
the transition depth $R_{T}$, is given by

\begin{equation}
B_{r}(\theta)=\sum_{n}(n+1)\left(\frac{R}{R_{T}}\right)^{n+2}g_{n}^{0}P_{n}(\sin\theta).\label{eq:Br}
\end{equation}

In the MFED balance it is assumed that the background field is known,
and the residual field induced by the flow is small in comparison
to the background. Saturn's measured magnetic field \citep{Dougherty2018},
given in terms of the Gauss coefficients $g_{i}^{0}$ (Table~\ref{tab:table1}),
can be separated into the main field, composed of $g_{1}^{0}$ through
$g_{3}^{0}$ (used as the background field), and the residual (potentially
wind-induced) field, composed of $g_{4}^{0}$ through $g_{11}^{0}$
(Figure~\ref{fig:Measurements}, red), potentially related to the
flow field. The separation into the main and residual fields stems
from the significant reduction in the value of $g_{4}^{0}$ compared
to $g_{3}^{0}$ (factor of $\sim25$), and might be attributed to
the existence of both a deep dynamo and an outer shallow dynamo \citep{Dougherty2018}.
A recent analysis of the Saturn magnetic field \citep{Cao2020} has
pointed to a very similar behavior, with similar conclusions regarding
the possibility that the residual field is induced by the flow in
the semiconducting region. Nonetheless, it can be argued that in the
newer analysis \citep{Cao2020} (in which $g_{12}^{0}$ to $g_{14}^{0}$
are also calculated), aside from $g_{3}^{0}$, $g_{4}^{0}$ and $g_{6}^{0}$,
the higher Gauss coefficients roughly correspond to a straight line
in the Lowes-Mauersberger power spectrum \citep{Lowes1974}, which
is an indication that higher harmonics are being generated by the
internal dynamo. As we will demonstrate, even if part of the residual
magnetic field is not induced by the flow, the results presented here
regarding the flow structure in the semiconducting region remain the
same. We therefore separate the measured gravity field into the main
field composed of $g_{1}^{0}$ to $g_{3}^{0}$, and the residual (potentially
wind-induced) field composed of $g_{4}^{0}$ to $g_{11}^{0}$, which
we attribute to the flow field. For simplicity we set the background
magnetic field $\mathbf{B_{0}}$ as function of $g_{1}^{0}$ only
(Equation~\ref{eq:Br}). Including $g_{2}^{0}$ and $g_{3}^{0}$
causes the wind-induced residual magnetic field to have a somewhat
smaller amplitude, but does not change qualitatively any of the results
reported here.

The constraint on the flow structure used to calculate the decay function
in the semiconducting region is taken as the magnitude of the magnetic
field $B_{r}$, calculated as the root mean square (RMS) defined between
60$^{\circ}$S and 60$^{\circ}$N
\begin{equation}
I=\sqrt{\frac{3}{2\pi}\intop_{-\pi/3}^{\pi/3}B_{r}(\theta)^{2}\,d\theta},\label{eq:I-RMS}
\end{equation}
being for the measured field $I=560\,$~nT for Saturn. Using the
MFED model we calculate $I$ for different combinations of the two
parameters defining the decay function $Q$ in the semiconducting
region: $0<f_{M}<1$ and $50\,{\rm km}<H_{M}<800{\rm \,km}$ (see
section~\ref{subsec:Definition-flow}).

In the MFED approximation all the non-axisymmetric dynamics are parameterized
for (using the dynamo $\alpha$-effect), and only the axisymmetric
magnetic field is solved for. This is an excellent assumption for
Saturn \citep{Dougherty2018,Garcia-Melendo2011}, while for Jupiter
where the magnetic field was found to exhibit strong east-west variations
\citep{Connerney2018} a method based on magnetic secular variations
is more appropriate \citep{Duer2019,Moore2019}. Another requirement
for using the MFED balance is that the magnetic Reynolds number would
satisfy $R_{m}(u)<1$ \citep{Cao2017a}. In Appendix~D we demonstrate
that this is indeed the case for Saturn and Jupiter. Few other factors
might affect the MFED solutions: First, the rotation period of Saturn
is still not fully known. In this study, we use the more recent estimates
of 10h 34m, but other rotation periods cannot be excluded. While it
was shown that the rotation period has very little effect on the wind-induced
gravity harmonics \citep{Galanti2017d}, it might have an affect on
the latitudinal variability of the residual magnetic field. Second,
the electrical conductivity used in this study is based on a limited
set of measurements and is estimated to have two orders of magnitude
uncertainty \citep{Liu2008}. In Appendix~C we discuss in detail
how this uncertainty affects our solutions. Finally, a more complex
$\alpha$-effect with latitudinal dependence, as well as the $\gamma$-effect
\citep{Kapyla2006}, are certainly possible and might be calculated
from 3D dynamo models, but there is currently uncertainty of what
values should be used for Saturn. However, while adding complexity
to the solutions, including these parameters should not change the
main results reported here.

\subsection{Definition of the flow structure\label{subsec:Definition-flow}}

In all variants of flow structure discussed in this study, the same
flow field is used to generate both the gravity and the magnetic fields.
We start by taking the observed meridional profile of wind at the
cloud-level (Figure~\ref{fig:Measurements}, black line, for Saturn,
and Figure~\ref{fig:Solutions-Jupiter}b, gray line, for Jupiter)
and decompose it into the first $N$ Legendre polynomials

\begin{eqnarray}
u_{{\rm }}^{{\rm obs}}(\theta) & = & \sum_{i=0}^{N}A_{i}^{{\rm obs}}P_{i}(\sin\theta),\label{eq:U-reconstruction}
\end{eqnarray}
where $A_{i}^{{\rm obs}}$ are the coefficients determining the latitudinal
shape of the observed wind, $\theta$ is the latitude, $P_{i}$ are
the Legendre polynomials, and $N=99$ is the number of polynomials
to be used. Defining a modified cloud-level wind

\begin{eqnarray}
u^{{\rm sol}}(\theta) & = & \sum_{i=0}^{N}A_{i}^{{\rm sol}}P_{i}(\sin\theta),\label{eq:U-reconstruction-1}
\end{eqnarray}
where $A_{i}^{{\rm sol}}$ are the modified coefficients, we allow
these coefficients to vary during the optimization process while making
sure they do not deviate considerably from their observed values.
Note that we construct the wind using a very large number of polynomials
to allow the wind solution to follow closely the observed wind. The
optimization procedure described in Appendix~A ensures that the large
number of coefficients is well constrained. Next, the modified cloud-level
wind $u^{{\rm sol}}(\theta)$ is projected parallel to the axis of
rotation \citep{Kaspi2009} to get the basic non-decaying field $u_{0}(r,\theta)$.
This field is then decayed in the radial direction to give
\begin{eqnarray}
u(r,\theta) & = & u_{{\rm 0}}(r,\theta)Q(r),\label{eq:MHD-flow-1}
\end{eqnarray}
where $r$ is the radial direction. The decay function $Q(r)$ is
defined as
\begin{equation}
Q(r)=\tanh\left(\frac{r-R_{T}}{\delta H_{T}}\right)\frac{1-f_{M}}{\tanh\left(\frac{R-R_{T}}{\delta H_{T}}\right)}+f_{M},\quad R_{T}<r<R,\label{eq:TW-decay-modified-1}
\end{equation}

\begin{equation}
Q(r)=f_{M}\exp\left(\frac{r-R_{T}}{H_{M}}\right),\quad r\leq R_{T},\label{eq:TW-decay-modified-inner-1}
\end{equation}
where $\delta H_{T}$ is the width of the hyperbolic tangent function,
$f_{M}$ is the ratio between the flow strength at the transition
depth and the flow at the cloud-level and $H_{M}$ is the decay scale-height
in the inner layer. This functional form of the flow's radial decay
allows two distinctly different behaviors in the regions above and
below the transition depth $R_{T}$. In the outer region the decay
function represents a non-magnetic dynamical effect, with the baroclinic
shear being in thermal wind balance \citep{Kaspi2009}, and the free
parameter $\delta H_{M}$ allowing a range of decaying profiles, from
a gradual decay to a case where the cloud-level winds keep their value
almost constant until reaching the transition depth. In the inner
region, the exponential decay function is assumed to be a result of
the increased electrical conductivity $\sigma$. Based on the choice
of the parameters $A_{i}^{{\rm sol}}$$,f_{M}$, $H_{M}$ and $\delta H_{T}$,
the flow structure is defined, and can be used to generate both the
wind-induced gravity field and the wind-induced magnetic field.

\section{Results}

The methodology presented above can be more readily applied to Saturn
given its highly axisymmetric magnetic field\citep{Dougherty2018,Cao2020},
while the magnetic field of Jupiter is highly non-axisymmetric \citep{Connerney2018,Moore2018},
making the usage of the MFED approximation much more challenging.
We therefore first investigate the Saturn case with the combined magnetic-gravity
analysis, and then discuss the Jupiter case where we substitute the
MFED method with insights gained from past measurements of the planet's
magnetic field secular variations \citep{Moore2019}.

\subsection{The Saturn case}

\begin{figure}
\begin{centering}
\includegraphics[scale=0.34]{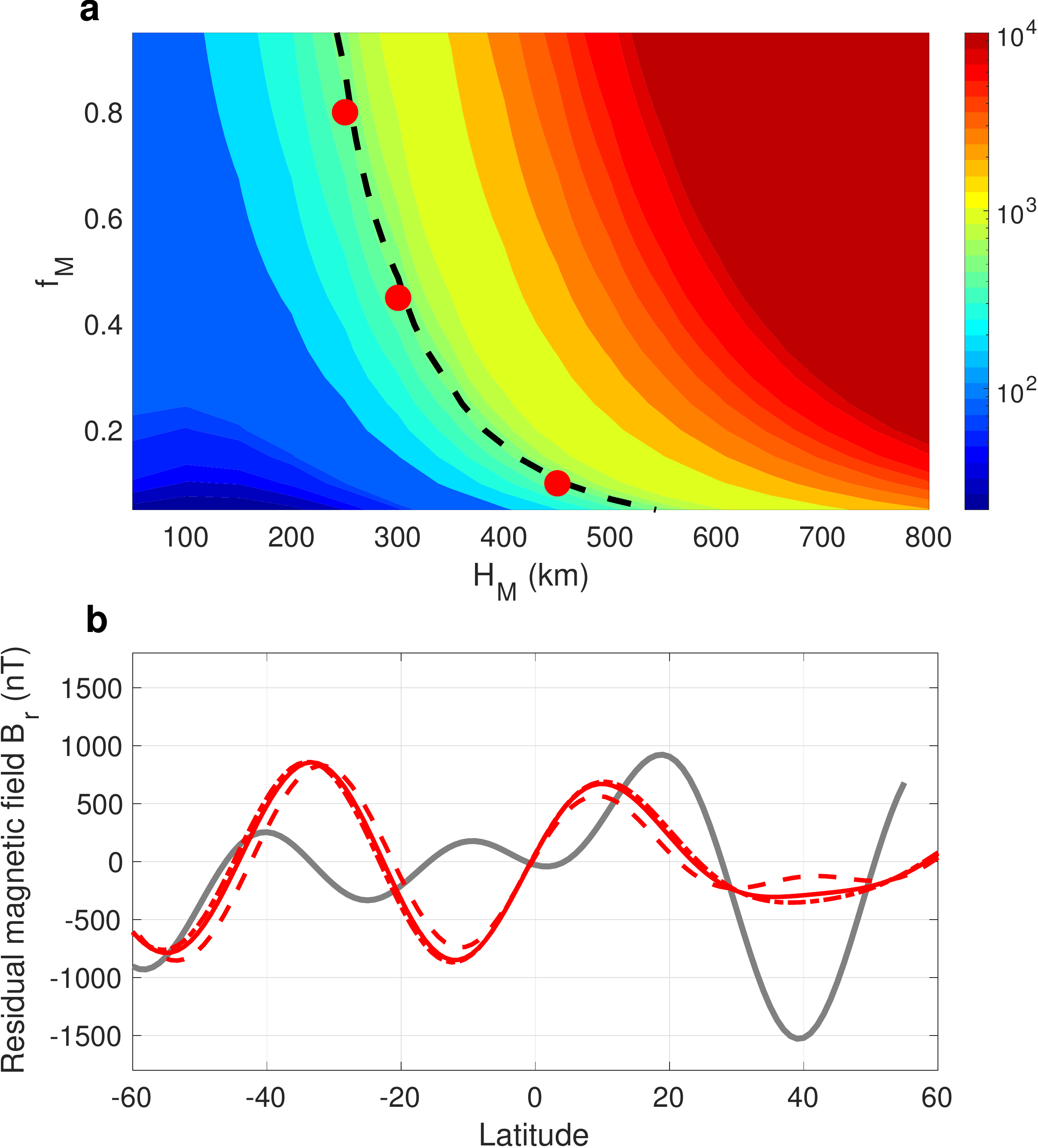}
\par\end{centering}
\centering{}\caption{\textbf{The residual magnetic field induced by the flow structure.}
(a) The RMS of the residual magnetic field $B_{r}$ at $0.875\,R_{s}$
for a range of flow decay parameters $H_{{\rm M}}$~(km) and $f_{{\rm M}}$.
Also shown is the value of the measured field of 560~nT (dashed contour)
and three representative combinations of parameters matching the measurements:
$H_{{\rm M}}=450,300,250\,{\rm km}$ and $f_{{\rm M}}=0.1,0.45,0.8$,
respectively. (b) The resulting latitude-dependent residual magnetic
field $B_{r}$ for the three representative combinations: dashed,
solid and dashed-doted, respectively. Also shown is the measured field
(gray).\label{fig:range_Hu_UM}}
\end{figure}

We start by using the Saturn magnetic field to constrain the flow
field. This constraint is set by the magnitude of the measured residual
magnetic field, defined here as the root mean square (RMS) of the
radial component, $B_{r}$, between 60$^{\circ}$S and 60$^{\circ}$N
(Figure~\ref{fig:Measurements}), calculated to be $I=560\,{\rm nT}$
(Equation~\ref{eq:I-RMS}). This is the maximal residual magnetic
signature expected from the flow field. By using the MFED balance
(section~\ref{subsec:MFED}) and varying the two parameters defining
the flow structure in the semiconducting region, $f_{{\rm M}}$ and
$H_{{\rm M}}$ (section~\ref{subsec:Definition-flow}), the full
range of possible solutions for the wind-induced magnetic field is
revealed (Figure~\ref{fig:range_Hu_UM}a). As expected, higher values
of $H_{{\rm M}}$ and $f_{{\rm M}}$ (deeper flows) give larger magnetic
signatures, with $H_{{\rm M}}$ being the dominant parameter. Since
the measured value of 560~nT can be obtained with different combinations
of the two parameters (Figure~\ref{fig:range_Hu_UM}a, dashed black
contour), the zonal wind's radial decay in the semiconducting region
cannot be determined uniquely. However, different parameter combinations
located along the 560~nT contour give very similar solutions for
$B_{r}$. For example, the three combinations denoted by the red dots
in Figure~\ref{fig:range_Hu_UM}a result in almost identical profiles
of $B_{r}$ (Figure~\ref{fig:range_Hu_UM}b, red lines). Most importantly,
the shape of the new decay profiles in the semiconducting region (Figure~\ref{fig:Solutions}a,
red lines deeper than the transition depth) are very different from
the solution constrained with only the gravity field \citep{Galanti2019a}
(dashed black line). The magnetic field measurements imply that the
flow must decay sharply at the transition depth. This depth is much
shallower than the depth where the gradually decaying gravity-only
solution loses most of the wind strength. In addition, due to the
magnetic field constraint, in the new solution the long tail does
not extend deep into the planet's interior (gray area).

\subsubsection{A magnetically-restricted solution\label{subsec:A-magnetically-restricted-solution}}

Having determined the wind decay functions in the semiconducting region
that are consistent with the magnetic field measurements (Table~\ref{tab:table2}),
an optimal solution is sought for the wind decay function above the
transition depth, such that the associated residual gravity field
matches the measured gravity field (Appendix~A). Using thermal wind
balance (section~\ref{subsec:TW}), the optimal decay function above
the transition depth (Figure~\ref{fig:Solutions}a, red lines) and
the optimal cloud-level wind structure (Figure~\ref{fig:Solutions}b,
red lines) are found, such that the resulting wind-induced gravity
field best explains the measured residual gravity field (Figure~\ref{fig:Measurements},
blue; see also values in Table~\ref{tab:table1}a). Note that since
the decay function is optimized only above the transition depth, the
solutions still satisfy the magnetic field constraint (Appendix~A).

\begin{table*}
\begin{centering}
\includegraphics[scale=0.95]{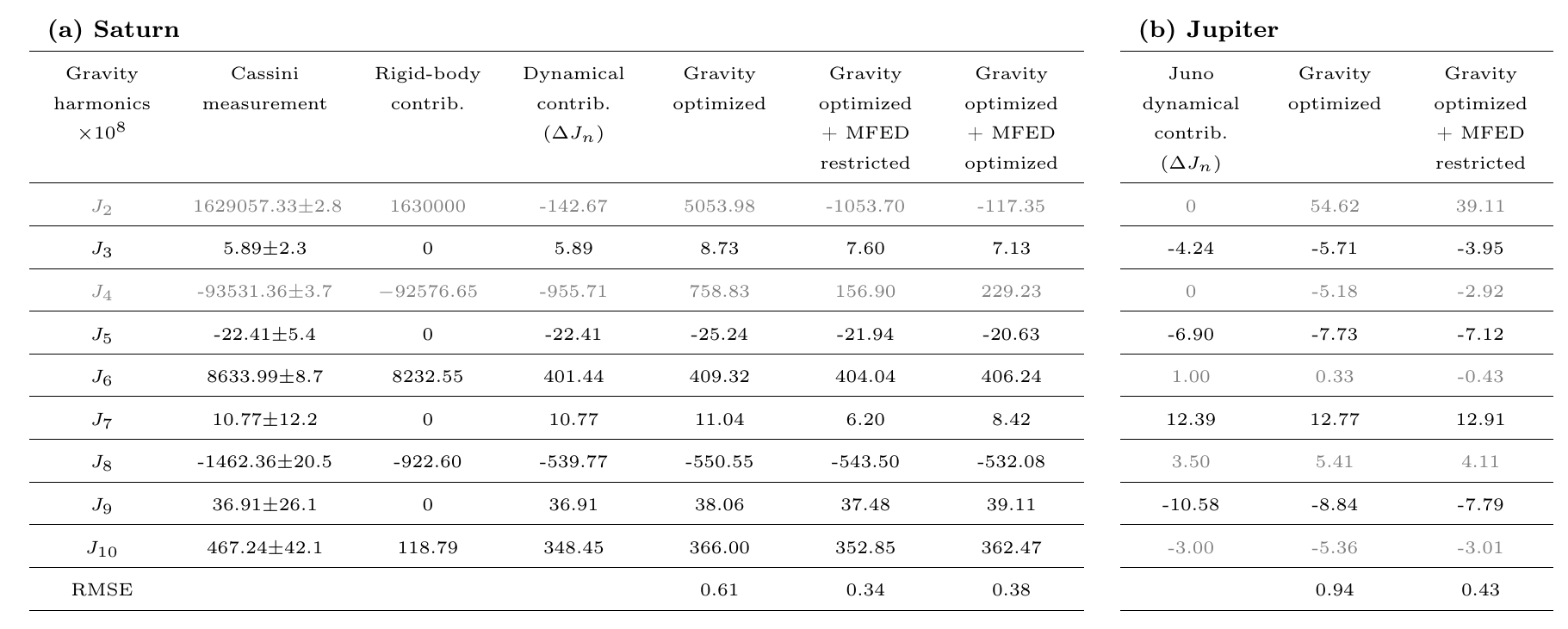}
\par\end{centering}\caption{\textbf{(a) The Saturn gravity harmonics $J_{2}$ to $J_{10}$.} Shown are the Cassini measurements \citep{Iess2019}, the rigid-body contribution
based on an average of ensemble model solutions \citep{Galanti2019a},
the dynamical contribution calculated as the difference between the
measurements and the rigid-body contribution, the gravity optimized
solution \citep{Galanti2019a} based on the gravity measurements only,
the new gravity-optimized and MFED-restricted solution, where the
MFED restricts the inner part of the decay profile but is not optimized
for, and the new gravity and MFED optimized solution where the cloud-level
wind is optimized to allow a fit for both gravity and magnetic fields.
Note that $J_{2}$ and $J_{4}$ (gray) are not optimized for. \textbf{(b)
The Jupiter gravity harmonics}. Shown are the dynamical contributions
(measured odds and estimated evens) \citep{Kaspi2020}, the solution
based only on the gravity measurements \citep{Kaspi2018}, and the
new gravity-optimized and MFED-restricted solution. Note that in the
Jupiter case, only $J_{3}$, $J_{5}$, $J_{7}$ and $J_{7}$ (black)
are optimized for. \label{tab:table1}}
\end{table*}

All three optimal solutions are within the measurement uncertainty,
with a root mean square error (Equation~\ref{eq:RMSE-grav}) of 0.34,
0.34 and 0.33 for the main decay profile (solid red), and the two
variants (dashed and dotted-dashed), respectively (Table~\ref{tab:table1}a,
note that an RMSE of 1 means that the solution harmonics are on average
at the measurement error distance from the measurement itself). Note
also that the new solutions derived by the magnetic field measurements
that dramatically restrict the flow to shallower depth, are not deteriorated
compared to the gravity-only solution (with an RMSE of 0.61). In fact,
they are better solutions in terms of the RMSE (Table~\ref{tab:table1}a).
This improvement is predominantly due to the different decay functions
used in the semiconducting region. While in the setup of the gravity-only
solution the possible decay function is a combination of synthetic
smoother functions, here the physically driven, much sharper decay
in the transition depth allows a better fit to the measurements. Moreover,
the required modification in the cloud-level wind (Figure~\ref{fig:Solutions}b,
solid red) is marginal with respect to the gravity-only solution,
strengthening the robustness of the wind solution with respect to
the observed wind. The shape of the new decay function in the outer
neutral region implies that, unlike in the gravity-only solution (Figure~\ref{fig:Solutions}a,
dashed black), the winds in the new solution barely decay in the first
7000~km below the cloud-level (i.e., barotropic); only below that
depth a decay commences. Only such strong flow in the outer region
can generate sufficient mass advection that would be able to explain
the gravity measurements.

\subsubsection{A fully combined gravity-magnetic solution}

\begin{figure}[t]
\centering{}\includegraphics[scale=0.35]{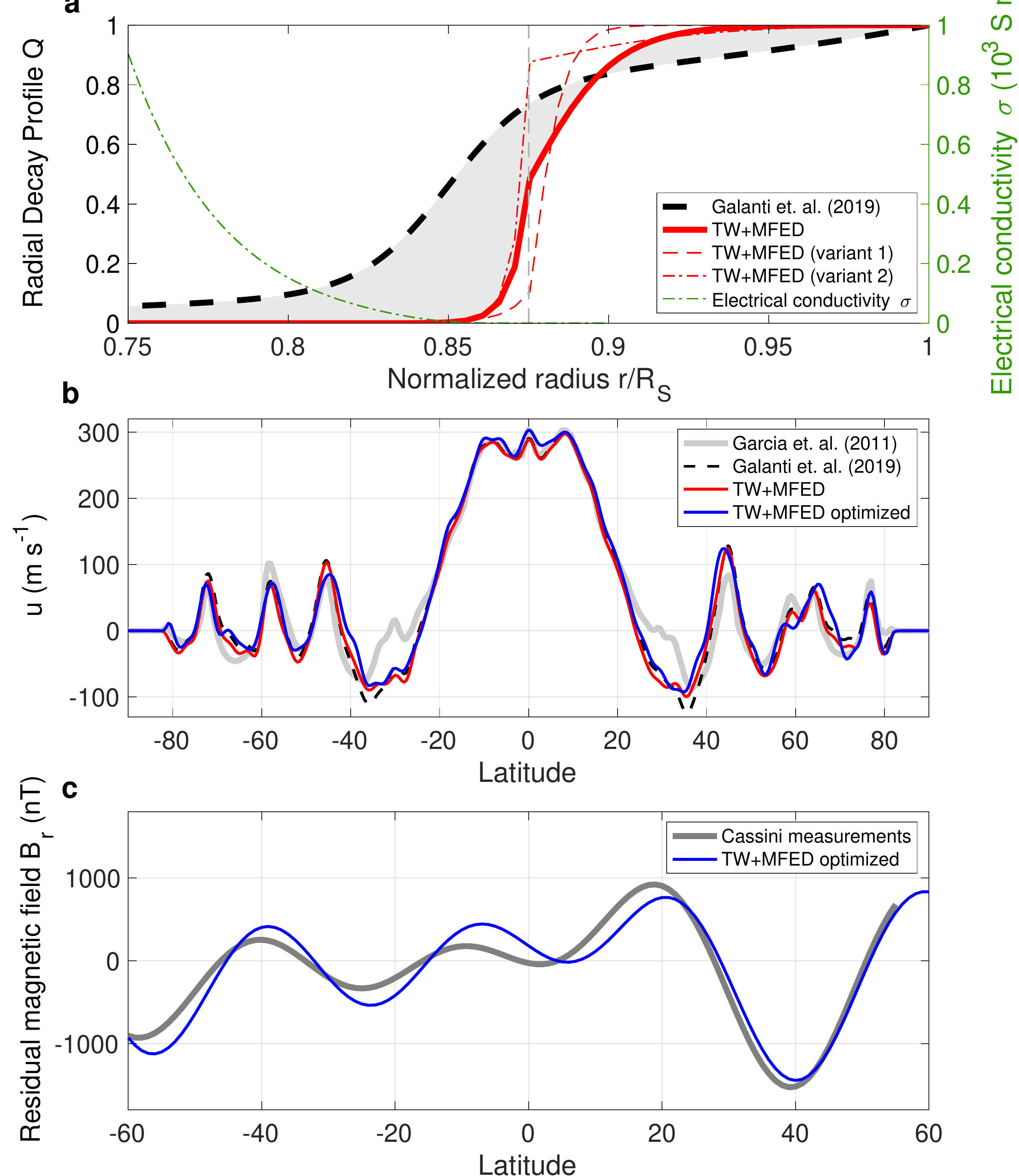}\caption{\textbf{The implication of the magnetic field constraint for Saturn's
flow structure. }(a) The new solution for the wind's radial decay
(solid red) together with the gravity-only constrained solution \citep{Galanti2019a}
(dashed black). The difference between the solutions is emphasized
by the gray area. Also shown are two representative variants of the
new solution corresponding to the red dots in Figure.~\ref{fig:range_Hu_UM}a
(dashed and dotted-dashed red), the conductivity profile $\sigma\:({\rm S\,m}{}^{-1}$,
green), and the transition depth ($0.875\,R_{S}$) separating the
outer neutral region and the inner semiconducting region (vertical
dashed line). (b) The cloud-level wind profiles. Shown are the observations
\citep{Garcia-Melendo2011} (gray), the new solution with the magnetic
field restricting the decay profile (red), the fully optimized new
solution with the residual magnetic field optimized in addition to
the gravity harmonics (blue), and the gravity-only restricted solution
\citep{Galanti2019a} (dashed black). (c) The magnetic field of the
fully optimized solution (blue) and the measured residual magnetic
field \citep{Dougherty2018} (gray).\label{fig:Solutions}}
\end{figure}

The new, depth-confined, solution fits all the relevant gravity harmonics
(Table~\ref{tab:table1}a) and the magnitude of the residual magnetic
field (Figure~\ref{fig:range_Hu_UM}b, solid red line, and Table~\ref{tab:table2}),
but its latitudinal dependence is different from the measured magnetic
field. By setting the wind decay profile to the solution already obtained
(Figure\ \ref{fig:Solutions}a, solid red line), we perform a full
optimization of the cloud-level wind, looking for a solution for which
the values of both the induced gravity and magnetic fields are within
the uncertainties of the measurements (Appendix~B). We obtain an
optimal solution (Figure\ \ref{fig:Solutions}b, blue line) with
an RMSE of 0.38 for the residual gravity field and an RMSE of 0.99
for the residual magnetic field (Figure\ \ref{fig:Solutions}c, Table~\ref{tab:table2}).
Therefore, with the same decay profile, and with only minor modifications
to the cloud-level wind, well within the measurement uncertainty \citep{Garcia-Melendo2011},
a flow field is found that can explain the latitudinal dependence
of \textit{both} the gravity and the magnetic fields. In summary,
we find that the values of the higher Gauss coefficients pose an upper
bound on the wind-induced magnetic field of Saturn. No matter whether
the residual field comes from the winds or the deep dynamo, the contributions
of the wind in the semiconducting region to the residual field cannot
be higher than the measurements, and this strongly constrains the
magnitude of the flow in the semiconducting region.

\subsubsection{Applicability of the solution}

The ability of the MFED balance to constraining the flow field in
the semiconducting region should be examined from several aspects.
First, a basic requirement for using the MFED balance is that the
magnetic Reynolds number, defined as $R_{m}(u)=uH_{\sigma}\sigma\mu_{0}$
($H_{\sigma}=\sigma/\frac{d\sigma}{dr}$ is the scale height associated
with it), would satisfy $R_{m}(u)<1$ \citep{Cao2017a}. We find that
$R_{m}(u)$ calculated with the optimal flow solution (Figure~\ref{fig:Solutions}a),
satisfies this condition everywhere (Appendix~D). Moreover, the wind-induced
residual magnetic field $B_{r}$ can be roughly related to the background
field $\mathbf{B_{0}}$ via $B_{r}\sim R_{m}(\alpha)^{{\rm max}}R_{m}(u)^{{\rm max}}\mathbf{|B_{0}|}$,
where $R_{m}(\alpha)=\alpha H_{\sigma}\sigma\mu_{0}$ is the magnetic
Reynolds number associated with the dynamo $\alpha$-effect \citep{Cao2017a}.
Since $R_{m}(\alpha)$ reaches a maximal value of 0.25 (at the lower
boundary of the semiconducting region) and $R_{m}(u)<1$ everywhere,
the resulting residual magnetic field is always much smaller than
the background field (Figure~\ref{fig:range_Hu_UM}b). Second, the
MFED balance depends on several parameters that are not strongly constrained.
Most importantly, the electrical conductivity $\sigma$ is known only
within 2 orders of magnitude \citep{Liu2008}. In order to evaluate
the effect of this uncertainty on our solution, we examine two extreme
cases, for which $\sigma$ is an order-of-magnitude larger and lower
than the mean value used here (Appendix~C). In both cases, solutions
for the flow field that satisfy both the magnetic and gravity measurements
can be found. In the former case, the solution shows an even more
pronounced barotropic behavior in the outer layers and an even sharper
decay of the winds near the transition depth. In the latter case,
the solution exhibits a more gradual wind decay, asymptotically getting
closer to (but still far from) the gravity-only solution (Figure~\ref{fig:range_Hu_UM-extremes-1}).

\begin{table}[t!]
\begin{centering}
\includegraphics[scale=0.5]{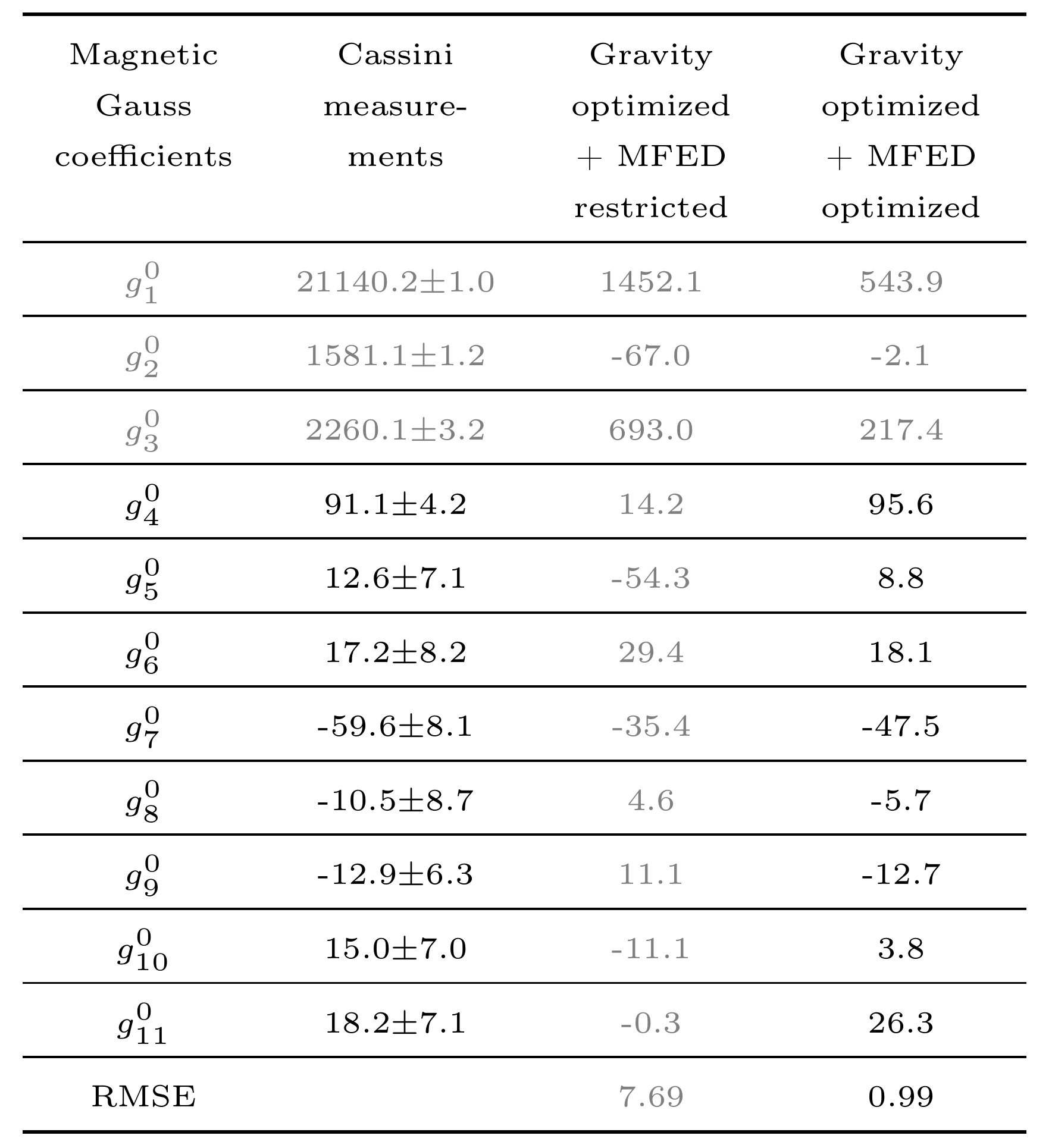}
\par\end{centering}\caption{\textbf{The Gauss coefficients composing the magnetic field of Saturn.} Shown are the Cassini measurements \citep{Dougherty2018}, the gravity-optimized
and MFED-restricted solution, and the optimized model solution corresponding
to column 7 in Table~\ref{tab:table1}. Note that in column 3 none
of the Gauss coefficients are optimized for, and in column 4, $g_{1}^{0}$,
$g_{2}^{0}$ and $g_{3}^{0}$ (gray) are not optimized for.\label{tab:table2}}
\end{table}

Moreover, these results also pose a limit on the uncertainty in the
conductivity profile, where conductivity that is 2-orders-of-magnitude
larger would not allow a physical solution, and that such a limit
does not exists for a lower value. Third, the dynamo $\alpha$-effect,
parameterizing the action of the small-scale 3D turbulence on the
wind induced toroidal magnetic field, is assumed in our study to be
only a function of depth, with a sign change at the equator \citep{Cao2017a}.
In the framework of the MFED balance, the uncertainty associated with
the magnitude of $\alpha$ is equivalent to an uncertainty in $\sigma$.
Therefore, its effect on our solutions is similar to the one discussed
in the analysis above. Finally, inhomogeneity of $\alpha$ in space
or time might also have an effect on the solution, yet these effects
are difficult to estimate, and might be compensated by small alterations
to the cloud-level wind.

\subsection{The Jupiter case}

The above results have direct implications for Jupiter, whose gravity
measurements were also used to decipher the flow structure \citep{Iess2018,Kaspi2018},
and whose optimal radial decay based on gravity alone was found to
be quite gradual. If the mechanisms governing the flow structure are
similar to Saturn's, constraining Jupiter's flow with the Juno magnetic
field measurements should show a sharp shear at a depth of around
2000~km below the cloud level. As the magnetic field of Jupiter is
highly non-axisymmetric \citep{Connerney2018,Moore2018}, using the
MFED approximation is much more challenging, since it involves a non-axisymmetric
wind-induced magnetic field, making it more difficult to disentangle
this field from the internal field. Recently, using a dynamo model,
the effect of different flow structures on the magnetic field in the
semiconducting region of Jupiter was estimated, yielding the conclusion
that the effect might be measurable by the Juno mission \citep{Wicht2019a}.
However, the Lowes spectrum of the higher magnetic harmonics of Jupiter
\citep{Connerney2018}, fits roughly a straight line, suggesting that
the internal field might play a dominant role in setting their values
\citep{Tsang2020}. Again, this renders the task of separating the
wind-induced magnetic field highly challenging.

\begin{figure}
\centering{}\includegraphics[scale=0.35]{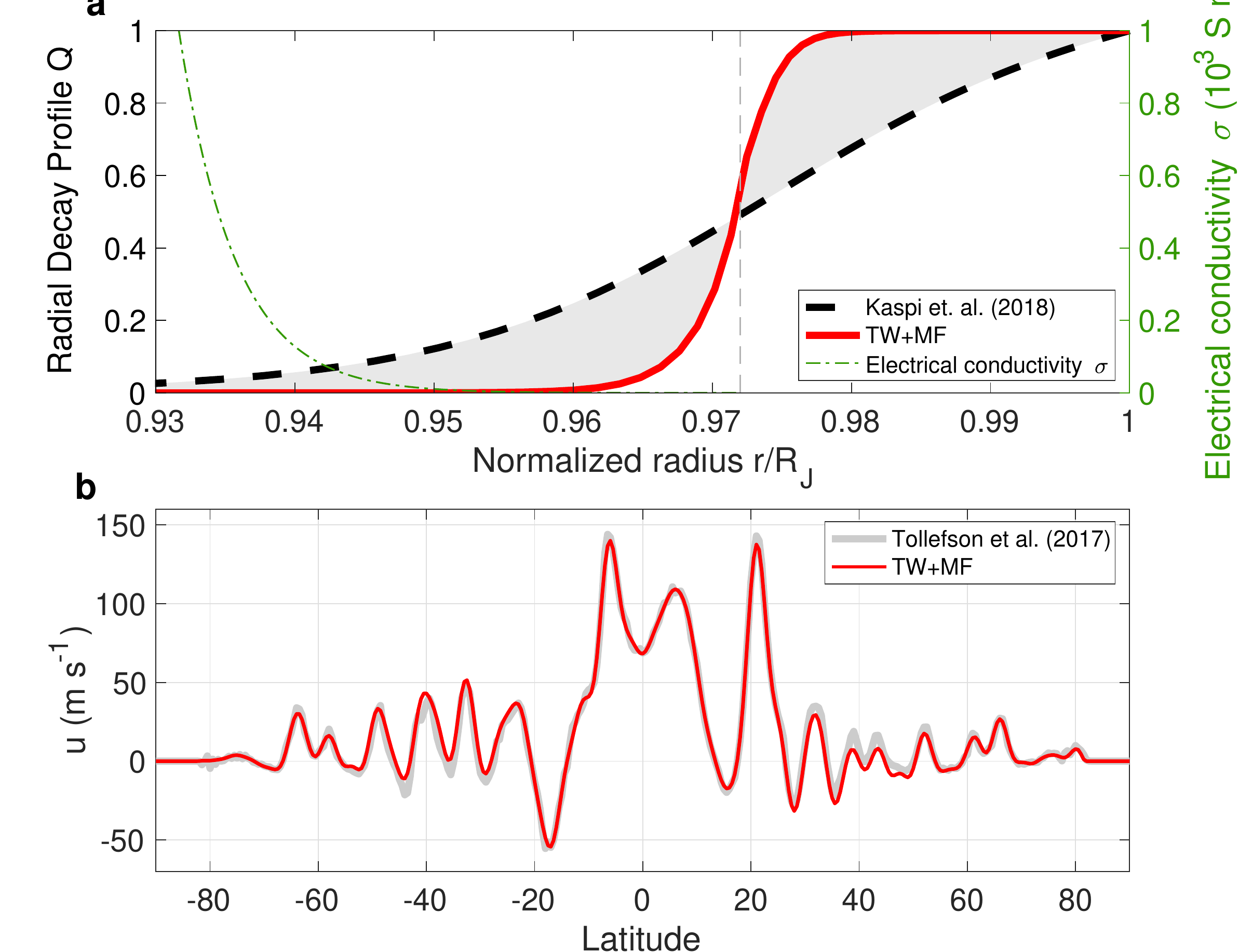}\caption{\textbf{The implication of the magnetic field constraint for Jupiter's
flow structure.} (a) The new solution for the wind's radial decay
(solid red) together with the gravity-only constrained solution \citep{Kaspi2018}
(dashed black). The difference between the solutions is emphasized
by the gray area. The decay function in the semi-conducting region
is defined with $H_{{\rm M}}=200\,{\rm km}$ and $f_{{\rm M}}=0.55$,
so that the wind is set to about $2\,{\rm cm\,s^{-1}}$ at $0.93\,R_{J}$
and $0.7\,{\rm m\,s^{-1}}$ at $0.95\,R_{J}$. In the outer region,
the decay function is defined with $\delta H_{T}=204$~km. Also shown is the conductivity
profile $\sigma\:({\rm S}\,{\rm m}^{-1},$ green) (see Supporting
Information for definition), and the transition depth ($0.972\,R_{J}$)
separating the outer neutral region and the inner semiconducting region
(vertical dashed line). (b) The cloud-level wind profiles. Shown are
the observations \citep{Tollefson2017} (gray) and the new solution
with the magnetic field restricting the decay profile (red).\label{fig:Solutions-Jupiter}}
\end{figure}

The effect of the wind on Jupiter's magnetic field might be identified
in the magnetic field's secular variations \citep{Ridley2016,Duer2019,Moore2019}.
Given that a time span of several years of magnetic field measurements
is required for calculating the secular variations, an analysis based
on the Juno measurements is expected to be available only toward the
end of the nominal mission \citep{Bolton2017}. However, we can already
use the available studies, and the results of this study, to examine
the potential implications for Jupiter. It has been shown, based on
the past measurements of Jupiter's magnetic field and their decadal
secular variation (SV), that the magnetic drift at a depth of $0.95\,R_{J}$
is of the order of a few centimeters per second \citep{Moore2019}.
The authors conclude that the flow itself can be restricted to the
values of the drift rate at $0.93\,R_{J}$, and that Ohmic dissipation
considerations limit the flow to an order of 1~m~${\rm s^{-1}}$
in the $0.94$ to $0.95\,R_{J}$ region \citep{Cao2017a}. Since the
gravity field is not very sensitive to this region, definitely not
to variations of the flow strength below 1~m~${\rm s^{-1}}$, we
can confidently choose a decay function that generates such a flow.

Assuming that the mechanism governing the structure of the flow is
similar to the one we find in Saturn, a flow structure in the semi-conducting
region ($0.95\,R_{J}$ to $0.972\,R_{J}$) can be set such that the
above constraint is met (Figure~\ref{fig:Solutions-Jupiter}a). The
magnetic Reynolds number associated with this solution ensures that
the wind-induced residual magnetic field is much smaller than the
internal field (Appendix~D). Similar to the case of Saturn, we search
for a solution for the flow structure in the outer layers ($r>0.972\,R_{J}$)
such that the measured odd gravity field is explained. By only slightly
varying the cloud-level winds (Figure~\ref{fig:Solutions-Jupiter}b),
within the measurement uncertainty \citep{Tollefson2017}, we find
a solution for the full decay function (Figure~\ref{fig:Solutions-Jupiter}a,
solid red line). The fit to the gravity field has an RMSE of 0.43,
an even better fit to the measurement than the 0.90 achieved with
the gravity field alone \citep{Kaspi2018} (see Table~\ref{tab:table1}b).
Importantly, the difference (gray area) between the gravity-only solution
and the new solution is substantial, even more than the one found
in the Saturn case. The new possible solution for Jupiter is of a
structure that is remarkably similar to the new solution for Saturn
(Figure~\ref{fig:Solutions}a, solid red), i.e., almost no decay
of the winds from the surface to a depth of around 1800~km, and then
a sharp decay over a depth of about 600~km. Similar to the Saturn
case, if future studies find the conductivity to be an order-of-magnitude
larger than the values used here \citep{Liu2008}, it will strengthen
the conclusion regarding the shape of the wind decay. However, an
even larger conductivity does not permit a physical solution. Conversely,
if the conductivity is found to be much weaker, then the magnetically
constrained solution will be more gradual, and closer to the one calculated
based only on the gravity field.

\section{Discussion and conclusion}

The inclusion of the Cassini magnetic measurements as an additional
constraint on Saturn's flow structure below the cloud-level unveils
a well-confined flow field that not only can explain the residual
magnetic field but also better explains the measured gravity field.
Based on constraints from the magnetic secular variation, a similar
structure is plausible also in Jupiter. The sharp baroclinic shear
of the flow in the semiconducting region, as well as the barotropic
structure of the flow from the cloud level down to the transition
depth, suggests that the flow interaction with the magnetic field
in the semiconducting region \citep{Liu2008,Cao2017a} plays an important
role in the wind's decay in the interior of Saturn, Jupiter, and potentially
other giant planets \citep{Kaspi2013c,Soyuer2020}. The barotropic
nature above this region implies that the observed momentum flux convergence
at the cloud-level of Jupiter \citep{Salyk2006} and Saturn \citep{DelGenio2007}
can drive the flow to great depths, perhaps by the downward control
principle \citep{Haynes1991,Liu2010}, until dissipation due to rising
conductivity and interaction with the magnetic field causes its almost
abrupt decay. The downward propagation mechanism was shown to be effective
also in the presence of a stable layer \citep{Showman2006}, which
might also act to decay the flow together with the action of the magnetic
field \citep{Christensen2020}. Such a stable layer was recently suggested
to be necessary for the case of Jupiter \citep{Debras2019}.

The results reported here are also in agreement with simulations of
the magnetohydrodynamics of gas giants \citep{Heimpel2005,Gastine2012,Heimpel2016,Duarte2018},
in which strong barotropic flows are found above the fully conducting
region, and much weaker baroclinic flows inside it. In the barotropic
region, the zonal flow is aligned with the axis of rotation according
to the Taylor-Proudman theorem (for the compressible case), and horizontal
entropy gradients must be small \citep{Jones2014}. This also implies
that the entropy expansion coefficient does not change considerably
with depth, as the decay rate of the flow is a product of the entropy
gradients and the entropy expansion coefficient \citep{Kaspi2009}.
The barotropicity of the neutral region indicates that there is no
significant mechanism acting to decay the flow until the semiconducting
region is reached. Compared to the gravity-only solution, the restriction
of the flow at depth by the magnetic field measurements, requires
the flow in the neutral region to be larger, in order for the mass
advection to be enough to explain the gravity measurements. We expect
the results reported in this study, together with the better understanding
of the internal structure of Saturn and Jupiter, to enable better
explaining the exact mechanism by which the winds on the gas giants
are generated, extended into the planet interior, and finally decay
at depth.

\textit{Acknowledgments:} The authors thank the Juno Interior Working Group for useful discussions.
This research was supported by the Israeli Space Agency and the Helen
Kimmel Center for Planetary Science at the Weizmann Institute.


\begin{thebibliography}{61}
\providecommand{\natexlab}[1]{#1}
\expandafter\ifx\csname urlstyle\endcsname\relax
  \providecommand{\doi}[1]{doi:\discretionary{}{}{}#1}\else
  \providecommand{\doi}{doi:\discretionary{}{}{}\begingroup
  \urlstyle{rm}\Url}\fi

\bibitem[{\textit{{Atkinson} et~al.}(1996)\textit{{Atkinson}, {Pollack}, and
  {Seiff}}}]{Atkinson1996}
{Atkinson}, D.~H., J.~B. {Pollack}, and A.~{Seiff}, {G}alileo doppler
  measurements of the deep zonal winds at {J}upiter, \textit{Science},
  \textit{272}, 842--843, 1996.

\bibitem[{\textit{{Bolton} et~al.}(2017)}]{Bolton2017}
{Bolton}, S.~J., et~al., Jupiter's interior and deep atmosphere: The initial
  pole-to-pole passes with the {J}uno spacecraft, \textit{Science},
  \textit{356}, 821--825, \doi{10.1126/science.aal2108}, 2017.

\bibitem[{\textit{{Cao} and {Stevenson}}(2017)}]{Cao2017a}
{Cao}, H., and D.~J. {Stevenson}, {Zonal flow magnetic field interaction in the
  semi-conducting region of giant planets}, \textit{Icarus}, \textit{296},
  59--72, \doi{10.1016/j.icarus.2017.05.015}, 2017.

\bibitem[{\textit{{Cao} and Stevenson}(2017)}]{Cao2017b}
{Cao}, H., and D.~J. Stevenson, Gravity and zonal flows of giant planets: From
  the {E}uler equation to the thermal wind equation, \textit{J. Geophys. Res.
  (Planets)}, \textit{122}, 686--700, \doi{10.1002/2017JE005272}, 2017.

\bibitem[{\textit{{Cao} et~al.}(2020)\textit{{Cao}, {Dougherty}, {Hunt},
  {Provan}, {Cowley}, {Bunce}, {Kellock}, and {Stevenson}}}]{Cao2020}
{Cao}, H., M.~K. {Dougherty}, G.~J. {Hunt}, G.~{Provan}, S.~W.~H. {Cowley},
  E.~J. {Bunce}, S.~{Kellock}, and D.~J. {Stevenson}, {The landscape of
  Saturn's internal magnetic field from the Cassini Grand Finale},
  \textit{Icarus}, \textit{344}, 113541, \doi{10.1016/j.icarus.2019.113541},
  2020.

\bibitem[{\textit{{Christensen} et~al.}(2020)\textit{{Christensen}, {Wicht},
  and {Dietrich}}}]{Christensen2020}
{Christensen}, U.~R., J.~{Wicht}, and W.~{Dietrich}, {Mechanisms for Limiting
  the Depth of Zonal Winds in the Gas Giant Planets}, \textit{Astrophys. J.},
  \textit{890}(1), 61, \doi{10.3847/1538-4357/ab698c}, 2020.

\bibitem[{\textit{{Connerney} et~al.}(2018)}]{Connerney2018}
{Connerney}, J.~E.~P., et~al., A new model of {J}upiter's magnetic field from
  {J}uno's first nine orbits, \textit{Geophys. Res. Lett.}, \textit{45}(6),
  2590--2596, \doi{10.1002/2018GL077312}, 2018.

\bibitem[{\textit{{Debras} and {Chabrier}}(2019)}]{Debras2019}
{Debras}, F., and G.~{Chabrier}, New models of {J}upiter in the context of
  {J}uno and {G}alileo, \textit{Astrophys. J.}, \textit{872}, 100--,
  \doi{10.3847/1538-4357/aaff65}, 2019.

\bibitem[{\textit{{Del Genio} et~al.}(2007)\textit{{Del Genio}, {Barbara},
  {Ferrier}, {Ingersoll}, {West}, {Vasavada}, {Spitale}, and
  {Porco}}}]{DelGenio2007}
{Del Genio}, A.~D., J.~M. {Barbara}, J.~{Ferrier}, A.~P. {Ingersoll}, R.~A.
  {West}, A.~R. {Vasavada}, J.~{Spitale}, and C.~C. {Porco}, Saturn eddy
  momentum fluxes and convection: First estimates from {C}assini images,
  \textit{Icarus}, \textit{189}(2), 479--492,
  \doi{10.1016/j.icarus.2007.02.013}, 2007.

\bibitem[{\textit{{Dougherty} et~al.}(2018)}]{Dougherty2018}
{Dougherty}, M.~K., et~al., Saturn's magnetic field revealed by the {C}assini
  {G}rand {F}inale, \textit{Science}, \textit{362}(6410), aat5434,
  \doi{10.1126/science.aat5434}, 2018.

\bibitem[{\textit{{Duarte} et~al.}(2018)\textit{{Duarte}, {Wicht}, and
  {Gastine}}}]{Duarte2018}
{Duarte}, L. D.~V., J.~{Wicht}, and T.~{Gastine}, {Physical conditions for
  {J}upiter-like dynamo models}, \textit{Icarus}, \textit{299}, 206--221,
  \doi{10.1016/j.icarus.2017.07.016}, 2018.

\bibitem[{\textit{{Duer} et~al.}(2019)\textit{{Duer}, {Galanti}, and
  {Kaspi}}}]{Duer2019}
{Duer}, K., E.~{Galanti}, and Y.~{Kaspi}, Analysis of {J}upiter's deep jets
  combining {J}uno gravity and time-varying magnetic field measurements,
  \textit{Astrophys. J. Let.}, \textit{879}(2), L22,
  \doi{10.3847/2041-8213/ab288e}, 2019.

\bibitem[{\textit{{Duer} et~al.}(2020)\textit{{Duer}, {Galanti}, and
  {Kaspi}}}]{Duer2020}
{Duer}, K., E.~{Galanti}, and Y.~{Kaspi}, {The Range of Jupiter's Flow
  Structures that Fit the Juno Asymmetric Gravity Measurements}, \textit{J.
  Geophys. Res. (Planets)}, \textit{125}(8), e06292,
  \doi{10.1029/2019JE006292}, 2020.

\bibitem[{\textit{{Fletcher} et~al.}(2020)}]{Fletcher2020}
{Fletcher}, L.~N., et~al., {Jupiter's Equatorial Plumes and Hot Spots: Spectral
  Mapping from Gemini/TEXES and Juno/MWR}, \textit{J. Geophys. Res. (Planets)},
  \textit{125}(8), e06399, \doi{10.1029/2020JE006399}, 2020.

\bibitem[{\textit{{French} et~al.}(2012)\textit{{French}, {Becker}, {Lorenzen},
  {Nettelmann}, {Bethkenhagen}, {Wicht}, and {Redmer}}}]{French2012}
{French}, M., A.~{Becker}, W.~{Lorenzen}, N.~{Nettelmann}, M.~{Bethkenhagen},
  J.~{Wicht}, and R.~{Redmer}, Ab initio simulations for material properties
  along the {J}upiter adiabat, \textit{Astrophys. J. Sup.}, \textit{202}(1), 5,
  \doi{10.1088/0067-0049/202/1/5}, 2012.

\bibitem[{\textit{{Galanti} and {Kaspi}}(2016)}]{Galanti2016}
{Galanti}, E., and Y.~{Kaspi}, {An Adjoint-based Method for the Inversion of
  the Juno and Cassini Gravity Measurements into Wind Fields},
  \textit{Astrophys. J.}, \textit{820}(2), 91,
  \doi{10.3847/0004-637X/820/2/91}, 2016.

\bibitem[{\textit{{Galanti} and {Kaspi}}(2017)}]{Galanti2017d}
{Galanti}, E., and Y.~{Kaspi}, {Prediction for the Flow-induced Gravity Field
  of Saturn: Implications for Cassini{\textquoteright}s Grand Finale},
  \textit{Astrophys. J. Let.}, \textit{843}(2), L25,
  \doi{10.3847/2041-8213/aa7aec}, 2017.

\bibitem[{\textit{{Galanti} et~al.}(2017{\natexlab{a}})\textit{{Galanti},
  {Cao}, and {Kaspi}}}]{Galanti2017e}
{Galanti}, E., H.~{Cao}, and Y.~{Kaspi}, Constraining {J}upiter's internal
  flows using {J}uno magnetic and gravity measurements, \textit{Geophys. Res.
  Lett.}, \textit{44}(16), 8173--8181, \doi{10.1002/2017GL074903},
  2017{\natexlab{a}}.

\bibitem[{\textit{{Galanti} et~al.}(2017{\natexlab{b}})\textit{{Galanti},
  {Kaspi}, and {Tziperman}}}]{Galanti2017a}
{Galanti}, E., Y.~{Kaspi}, and E.~{Tziperman}, {A full, self-consistent
  treatment of thermal wind balance on oblate fluid planets}, \textit{J. Fluid
  Mech.}, \textit{810}, 175--195, \doi{10.1017/jfm.2016.687},
  2017{\natexlab{b}}.

\bibitem[{\textit{{Galanti} et~al.}(2019)\textit{{Galanti}, {Kaspi}, {Miguel},
  {Guillot}, {Durante}, {Racioppa}, and {Iess}}}]{Galanti2019a}
{Galanti}, E., Y.~{Kaspi}, Y.~{Miguel}, T.~{Guillot}, D.~{Durante},
  P.~{Racioppa}, and L.~{Iess}, Saturn's deep atmospheric flows revealed by the
  {C}assini grand finale gravity measurements, \textit{Geophys. Res. Lett.},
  \textit{46}(2), 616--624, \doi{10.1029/2018GL078087}, 2019.

\bibitem[{\textit{{Garc{\'\i}a-Melendo}
  et~al.}(2011)\textit{{Garc{\'\i}a-Melendo}, {P{\'e}rez-Hoyos},
  {S{\'a}nchez-Lavega}, and {Hueso}}}]{Garcia-Melendo2011}
{Garc{\'\i}a-Melendo}, E., S.~{P{\'e}rez-Hoyos}, A.~{S{\'a}nchez-Lavega}, and
  R.~{Hueso}, Saturn's zonal wind profile in 2004-2009 from {C}assini {ISS}
  images and its long-term variability, \textit{Icarus}, \textit{215}(1),
  62--74, \doi{10.1016/j.icarus.2011.07.005}, 2011.

\bibitem[{\textit{{Gastine} and {Wicht}}(2012)}]{Gastine2012}
{Gastine}, T., and J.~{Wicht}, Effects of compressibility on driving zonal flow
  in gas giants, \textit{Icarus}, \textit{219}, 428--442,
  \doi{10.1016/j.icarus.2012.03.018}, 2012.

\bibitem[{\textit{{Gastine} et~al.}(2014)\textit{{Gastine}, {Wicht}, {Duarte},
  {Heimpel}, and {Becker}}}]{Gastine2014}
{Gastine}, T., J.~{Wicht}, L.~D.~V. {Duarte}, M.~{Heimpel}, and A.~{Becker},
  Explaining {J}upiter's magnetic field and equatorial jet dynamics,
  \textit{Geophys. Res. Lett.}, \textit{41}, 5410--5419,
  \doi{10.1002/2014GL060814}, 2014.

\bibitem[{\textit{{Haynes} et~al.}(1991)\textit{{Haynes}, {McIntyre},
  {Shepherd}, {Marks}, and {Shine}}}]{Haynes1991}
{Haynes}, P.~H., M.~E. {McIntyre}, T.~G. {Shepherd}, C.~J. {Marks}, and K.~P.
  {Shine}, {On the `Downward Control' of Extratropical Diabatic Circulations by
  Eddy-Induced Mean Zonal Forces.}, \textit{J. Atmos. Sci.}, \textit{48}(4),
  651--680, \doi{10.1175/1520-0469(1991)048<0651:OTCOED>2.0.CO;2}, 1991.

\bibitem[{\textit{{Heimpel} et~al.}(2005)\textit{{Heimpel}, {Aurnou}, and
  {Wicht}}}]{Heimpel2005}
{Heimpel}, M., J.~{Aurnou}, and J.~{Wicht}, {Simulation of equatorial and
  high-latitude jets on Jupiter in a deep convection model}, \textit{Nature},
  \textit{438}(7065), 193--196, \doi{10.1038/nature04208}, 2005.

\bibitem[{\textit{{Heimpel} et~al.}(2016)\textit{{Heimpel}, {Gastine}, and
  {Wicht}}}]{Heimpel2016}
{Heimpel}, M., T.~{Gastine}, and J.~{Wicht}, Simulation of deep-seated zonal
  jets and shallow vortices in gas giant atmospheres, \textit{Nature
  Geoscience}, \textit{9}, 19--23, \doi{10.1038/ngeo2601}, 2016.

\bibitem[{\textit{{Helled} et~al.}(2015)\textit{{Helled}, {Galanti}, and
  {Kaspi}}}]{Helled2015}
{Helled}, R., E.~{Galanti}, and Y.~{Kaspi}, {Saturn's fast spin determined from
  its gravitational field and oblateness}, \textit{Nature}, \textit{520}(7546),
  202--204, \doi{10.1038/nature14278}, 2015.

\bibitem[{\textit{Holton}(1992)}]{Holton1992}
Holton, J.~R., \textit{An Introduction to Dynamic Meteorology}, third ed., 511
  pp., Academic Press, 1992.

\bibitem[{\textit{{Hubbard}}(1999)}]{Hubbard1999}
{Hubbard}, W.~B., Note: Gravitational signature of {J}upiter's deep zonal
  flows, \textit{Icarus}, \textit{137}, 357--359, \doi{10.1006/icar.1998.6064},
  1999.

\bibitem[{\textit{{Iess} et~al.}(2018)}]{Iess2018}
{Iess}, L., et~al., Measurement of {J}upiter's asymmetric gravity field,
  \textit{Nature}, \textit{555}, 220--222, \doi{10.1038/nature25776}, 2018.

\bibitem[{\textit{{Iess} et~al.}(2019)}]{Iess2019}
{Iess}, L., et~al., Measurement and implications of {S}aturn's gravity field
  and ring mass, \textit{Science}, \textit{364}, 1052--, 2019.

\bibitem[{\textit{{Jones}}(2014)}]{Jones2014}
{Jones}, C.~A., A dynamo model of {J}upiter's magnetic field, \textit{Icarus},
  \textit{241}, 148--159, \doi{10.1016/j.icarus.2014.06.020}, 2014.

\bibitem[{\textit{Kapyla et~al.}(2006)\textit{Kapyla, Korpi, Ossendrijver, and
  Stix}}]{Kapyla2006}
Kapyla, P.~J., M.~J. Korpi, M.~Ossendrijver, and M.~Stix, Magnetoconvection and
  dynamo coefficients - iii. $\alpha$-effect and magnetic pumping in the rapid
  rotation regime, \textit{Astron. and Astrophys.}, \textit{455}(2), 401--412,
  \doi{10.1051/0004-6361:20064972}, 2006.

\bibitem[{\textit{{Kaspi}}(2013)}]{Kaspi2013a}
{Kaspi}, Y., Inferring the depth of the zonal jets on {J}upiter and {S}aturn
  from odd gravity harmonics, \textit{Geophys. Res. Lett.}, \textit{40},
  676--680, \doi{10.1029/2009GL041385}, 2013.

\bibitem[{\textit{{Kaspi} et~al.}(2009)\textit{{Kaspi}, {Flierl}, and
  {Showman}}}]{Kaspi2009}
{Kaspi}, Y., G.~R. {Flierl}, and A.~P. {Showman}, {The deep wind structure of
  the giant planets: Results from an anelastic general circulation model},
  \textit{Icarus}, \textit{202}(2), 525--542,
  \doi{10.1016/j.icarus.2009.03.026}, 2009.

\bibitem[{\textit{{Kaspi} et~al.}(2010)\textit{{Kaspi}, {Hubbard}, {Showman},
  and {Flierl}}}]{Kaspi2010a}
{Kaspi}, Y., W.~B. {Hubbard}, A.~P. {Showman}, and G.~R. {Flierl},
  Gravitational signature of {J}upiter's internal dynamics, \textit{Geophys.
  Res. Lett.}, \textit{37}, L01,204, \doi{10.1029/2009GL041385}, 2010.

\bibitem[{\textit{{Kaspi} et~al.}(2013)\textit{{Kaspi}, {Showman}, {Hubbard},
  {Aharonson}, and {Helled}}}]{Kaspi2013c}
{Kaspi}, Y., A.~P. {Showman}, W.~B. {Hubbard}, O.~{Aharonson}, and R.~{Helled},
  Atmospheric confinement of jet-streams on {U}ranus and {N}eptune,
  \textit{Nature}, \textit{497}, 344--347, \doi{10.1029/2009GL041385}, 2013.

\bibitem[{\textit{{Kaspi} et~al.}(2020)\textit{{Kaspi}, {Galanti}, {Showman},
  {Stevenson}, {Guillot}, {Iess}, and {Bolton}}}]{Kaspi2020}
{Kaspi}, Y., E.~{Galanti}, A.~P. {Showman}, D.~J. {Stevenson}, T.~{Guillot},
  L.~{Iess}, and S.~J. {Bolton}, {Comparison of the Deep Atmospheric Dynamics
  of Jupiter and Saturn in Light of the Juno and Cassini Gravity Measurements},
  \textit{Space Sci. Rev.}, \textit{216}(5), 84,
  \doi{10.1007/s11214-020-00705-7}, 2020.

\bibitem[{\textit{{Kaspi} et~al.}(2018)}]{Kaspi2018}
{Kaspi}, Y., et~al., Jupiter's atmospheric jet streams extend thousands of
  kilometres deep, \textit{Nature}, \textit{555}, 223--226,
  \doi{10.1038/nature25793}, 2018.

\bibitem[{\textit{{Kong} et~al.}(2018)\textit{{Kong}, {Zhang}, {Schubert}, and
  {Anderson}}}]{Kong2018}
{Kong}, D., K.~{Zhang}, G.~{Schubert}, and J.~D. {Anderson}, Origin of
  {J}upiter's cloud-level zonal winds remains a puzzle even after {J}uno,
  \textit{Proc. Natl. Acad. Sci. U.S.A.}, \textit{115}(34), 8499--8504,
  \doi{10.1073/pnas.1805927115}, 2018.

\bibitem[{\textit{{Liu} and {Schneider}}(2010)}]{Liu2010}
{Liu}, J., and T.~{Schneider}, {Mechanisms of Jet Formation on the Giant
  Planets}, \textit{Journal of Atmospheric Sciences}, \textit{67}(11),
  3652--3672, \doi{10.1175/2010JAS3492.1}, 2010.

\bibitem[{\textit{{Liu} et~al.}(2008)\textit{{Liu}, {Goldreich}, and
  {Stevenson}}}]{Liu2008}
{Liu}, J., P.~M. {Goldreich}, and D.~J. {Stevenson}, Constraints on deep-seated
  zonal winds inside {J}upiter and {S}aturn, \textit{Icarus}, \textit{196},
  653--664, \doi{10.1016/j.icarus.2007.11.036}, 2008.

\bibitem[{\textit{Lowes}(1974)}]{Lowes1974}
Lowes, F.~J., {Spatial Power Spectrum of the Main Geomagnetic Field, and
  Extrapolation to the Core}, \textit{Geophysical Journal International},
  \textit{36}(3), 717--730, \doi{10.1111/j.1365-246X.1974.tb00622.x}, 1974.

\bibitem[{\textit{{Mankovich} et~al.}(2019)\textit{{Mankovich}, {Marley},
  {Fortney}, and {Movshovitz}}}]{Mankovich2019}
{Mankovich}, C., M.~S. {Marley}, J.~J. {Fortney}, and N.~{Movshovitz}, Cassini
  ring seismology as a probe of {S}aturn's interior. {I}. rigid rotation,
  \textit{Astrophys. J.}, \textit{871}, 1, \doi{10.3847/1538-4357/aaf798},
  2019.

\bibitem[{\textit{{Moore} et~al.}(2018)}]{Moore2018}
{Moore}, K., et~al., A complex {J}ovian dynamo from the hemispheric dichotomy
  of {J}upiter's field, \textit{Nature}, \textit{561}, 76--78, 2018.

\bibitem[{\textit{{Moore} et~al.}(2019)\textit{{Moore}, {Cao}, {Bloxham},
  {Stevenson}, {Connerney}, and {Bolton}}}]{Moore2019}
{Moore}, K.~M., H.~{Cao}, J.~{Bloxham}, D.~J. {Stevenson}, J.~E.~P.
  {Connerney}, and S.~J. {Bolton}, {Time variation of Jupiter's internal
  magnetic field consistent with zonal wind advection}, \textit{Nature
  Astronomy}, \textit{3}, 730--735, \doi{10.1038/s41550-019-0772-5}, 2019.

\bibitem[{\textit{Pedlosky}(1987)}]{Pedlosky1987}
Pedlosky, J., \textit{Geophysical Fluid Dynamics}, pp.~710.~Springer-Verlag,
  1987.

\bibitem[{\textit{{Porco} et~al.}(2003)}]{Porco2003}
{Porco}, C.~C., et~al., {Cassini Imaging of Jupiter's Atmosphere, Satellites,
  and Rings}, \textit{Science}, \textit{299}(5612), 1541--1547,
  \doi{10.1126/science.1079462}, 2003.

\bibitem[{\textit{{Qin} et~al.}(2020)\textit{{Qin}, {Kong}, {Zhang},
  {Schubert}, and {Huang}}}]{Qin2020}
{Qin}, S., D.~{Kong}, K.~{Zhang}, G.~{Schubert}, and Y.~{Huang}, {Interpreting
  the Equatorially Antisymmetric Gravitational Field of Saturn Measured by the
  Cassini Grand Finale}, \textit{Astrophys. J.}, \textit{890}(1), 26,
  \doi{10.3847/1538-4357/ab6a9a}, 2020.

\bibitem[{\textit{{Read} et~al.}(2009)\textit{{Read}, {Dowling}, and
  {Schubert}}}]{Read2009}
{Read}, P.~L., T.~E. {Dowling}, and G.~{Schubert}, {S}aturn's rotation period
  from its atmospheric planetary-wave configuration, \textit{Nature},
  \textit{460}, 608--610, \doi{10.1038/nature08194}, 2009.

\bibitem[{\textit{{Ridley} and {Holme}}(2016)}]{Ridley2016}
{Ridley}, V.~A., and R.~{Holme}, Modeling the {J}ovian magnetic field and its
  secular variation using all available magnetic field observations, \textit{J.
  Geophys. Res. (Planets)}, \textit{121}(3), 309--337,
  \doi{10.1002/2015JE004951}, 2016.

\bibitem[{\textit{{Salyk} et~al.}(2006)\textit{{Salyk}, {Ingersoll}, {Lorre},
  {Vasavada}, and {Del Genio}}}]{Salyk2006}
{Salyk}, C., A.~P. {Ingersoll}, J.~{Lorre}, A.~{Vasavada}, and A.~D. {Del
  Genio}, Interaction between eddies and mean flow in {J}upiter's atmosphere:
  Analysis of {C}assini imaging data, \textit{Icarus}, \textit{185}, 430--442,
  \doi{10.1016/j.icarus.2006.08.007}, 2006.

\bibitem[{\textit{{S{\'a}nchez-Lavega}
  et~al.}(2000)\textit{{S{\'a}nchez-Lavega}, {Rojas}, and
  {Sada}}}]{Sanchez-Lavega2000}
{S{\'a}nchez-Lavega}, A., J.~F. {Rojas}, and P.~V. {Sada}, Saturn's zonal winds
  at cloud level, \textit{Icarus}, \textit{147}, 405--420,
  \doi{10.1006/icar.2000.6449}, 2000.

\bibitem[{\textit{{Showman} et~al.}(2006)\textit{{Showman}, {Gierasch}, and
  {Lian}}}]{Showman2006}
{Showman}, A.~P., P.~J. {Gierasch}, and Y.~{Lian}, Deep zonal winds can result
  from shallow driving in a giant-planet atmosphere, \textit{Icarus},
  \textit{182}, 513--526, \doi{10.1016/j.icarus.2006.01.019}, 2006.

\bibitem[{\textit{{Smith} et~al.}(1982)}]{Smith1982}
{Smith}, B.~A., et~al., A new look at the {S}aturn system: The {V}oyager 2
  images, \textit{Science}, \textit{215}, 505--537, 1982.

\bibitem[{\textit{{Soyuer} et~al.}(2020)\textit{{Soyuer}, {Soubiran}, and
  {Helled}}}]{Soyuer2020}
{Soyuer}, D., F.~{Soubiran}, and R.~{Helled}, {Constraining the depth of the
  winds on Uranus and Neptune via Ohmic dissipation}, \textit{MNRAS},
  \textit{498}(1), 621--638, \doi{10.1093/mnras/staa2461}, 2020.

\bibitem[{\textit{Tollefson et~al.}(2017)}]{Tollefson2017}
Tollefson, J., et~al., Changes in {J}upiter's zonal wind profile preceding and
  during the {J}uno mission, \textit{Icarus}, \textit{296}, 163--178, 2017.

\bibitem[{\textit{Tsang and Jones}(2020)}]{Tsang2020}
Tsang, Y.-K., and C.~A. Jones, Characterising {J}upiter's dynamo radius using
  its magnetic energy spectrum, \textit{Earth Planet. Sci. Lett.},
  \textit{530}, 115,879, \doi{https://doi.org/10.1016/j.epsl.2019.115879},
  2020.

\bibitem[{\textit{{Wicht} et~al.}(2019{\natexlab{a}})\textit{{Wicht},
  {Gastine}, and {Duarte}}}]{Wicht2019b}
{Wicht}, J., T.~{Gastine}, and L.~D.~V. {Duarte}, {Dynamo Action in the Steeply
  Decaying Conductivity Region of Jupiter-Like Dynamo Models}, \textit{J.
  Geophys. Res. (Planets)}, \textit{124}(3), 837--863,
  \doi{10.1029/2018JE005759}, 2019{\natexlab{a}}.

\bibitem[{\textit{{Wicht} et~al.}(2019{\natexlab{b}})\textit{{Wicht},
  {Gastine}, {Duarte}, and {Dietrich}}}]{Wicht2019a}
{Wicht}, J., T.~{Gastine}, L.~D.~V. {Duarte}, and W.~{Dietrich}, {Dynamo action
  of the zonal winds in {J}upiter}, \textit{Astron. and Astrophys.},
  \textit{629}, A125, \doi{10.1051/0004-6361/201935682}, 2019{\natexlab{b}}.

\bibitem[{\textit{{Zhang} et~al.}(2015)\textit{{Zhang}, {Kong}, and
  {Schubert}}}]{Zhang2015}
{Zhang}, K., D.~{Kong}, and G.~{Schubert}, {Thermal-gravitational Wind Equation
  for the Wind-induced Gravitational Signature of Giant Gaseous Planets:
  Mathematical Derivation, Numerical Method, and Illustrative Solutions},
  \textit{Astrophys. J.}, \textit{806}(2), 270,
  \doi{10.1088/0004-637X/806/2/270}, 2015.

\end{thebibliography}

\appendix

\section{A magnetically-restricted gravity optimization method\label{sec:App-solve-grav}}

The first set of solutions are performed in the following way. First,
the magnetic measurements are used to constrain the decay profile
in the semiconducting region. Then the thermal wind model is used
to find the optimal decay function in the outer non-conducting region,
as well as the optimal cloud-level wind, so that the resulting gravity
field explains best the gravity measurements. For the Saturn case,
the values used as measurements are the Cassini measurements \citep{Iess2019}
from which the static body harmonics \citep{Galanti2019a} are subtracted
\begin{equation}
J_{n}^{{\rm dyn}}=J_{n}^{{\rm obs}}-J_{n}^{{\rm rigid}},\label{eq:obs-TW}
\end{equation}
where $J_{n}^{{\rm obs}}$ are the measured gravity harmonics, and
$J_{n}^{{\rm rigid}}$ are the rigid-body solutions taken from the
average of an ensemble model solutions \citep{Galanti2019a}. Note
that $J_{n}^{{\rm rigid}}$ have non-zero values only for the even
harmonics. For the Jupiter case, only the odd gravity measurements
are used \citep{Iess2018}, and since these are fully wind-induced,
there is no need to subtract the solid body contribution \citep{Kaspi2018}
. 

The parameters to be optimized, i.e., the parameter defining the flow
structure above the semiconducting region and the cloud-level wind
latitudinal profile, are defined as a control vector
\begin{eqnarray}
\mathbf{X}_{{\rm C}} & = & \{\mathbf{X}_{H},\mathbf{X}_{u}\}\label{eq:control_vector-augmented}\\
 & = & \left\{ \delta H_{T}/h_{{\rm nor}},\left[A_{1}^{{\rm sol}},\cdots,A_{N}^{{\rm sol}}\right]/u_{{\rm nor}}\right\} ,\nonumber 
\end{eqnarray}
where $h_{{\bf nor}}=10^{7}$ and $u_{{\bf nor}}=10^{3}$ are the
normalization factors for the decay structure and the wind coefficients,
respectively. The normalization factors are chosen so that $0<\delta H_{T}/h_{{\rm nor}}<1$
and $-1<A^{{\rm {\rm sol}}}/u_{{\rm nor}}<1$ . Minimizing the difference
between the model solution for the gravity field and the measurements,
subjected to the uncertainties of the measurements and the need to
keep the optimized control parameters regularized to physical values,
is achieved with the cost function
\begin{eqnarray}
L_{T} & = & \mathbf{{\displaystyle \mathbf{\left(J^{m}-\mathbf{J}^{o}\right)}W\left(\mathbf{J^{m}}-\mathbf{J}^{o}\right)^{T}}}\label{eq:cost_function-augmented}\\
 & + & \epsilon_{u}\mathbf{(X_{u}-\mathbf{X}_{o})(X_{u}-\mathbf{X}_{o})^{T}},\nonumber 
\end{eqnarray}
where for the Saturn case $\mathbf{J^{m}}=\left[J_{3}^{m},J_{5}^{m},J_{6}^{m},J_{7}^{m},J_{8}^{m},J_{9}^{m},J_{10}^{m}\right]$
and $\mathbf{J^{o}}=\left[J_{3}^{{\rm dyn}},J_{5}^{{\rm dyn}},J_{6}^{{\rm dyn}},J_{7}^{{\rm dyn}},J_{8}^{{\rm dyn}},J_{9}^{{\rm dyn}},J_{10}^{{\rm dyn}}\right]$
are the calculated and measured gravity harmonics, respectively, ${\bf W}$
are the uncertainties in the gravity measurements, $\mathbf{X}_{{\rm o}}=\left[A_{1}^{{\rm obs}},\cdots,A_{99}^{{\rm obs}}\right]/u_{{\rm nor}}$
are the observed wind profile parameters and $\epsilon_{u}=5\times10^{8}$
is the weight given to the regularization of the wind solution to
the observed one. For the Jupiter case, $\mathbf{J^{m}}=\left[J_{3}^{m},J_{5}^{m},J_{7}^{m},J_{9}^{m}\right]$
and $\mathbf{J^{o}}=\left[J_{3}^{{\rm dyn}},J_{5}^{{\rm dyn}},J_{7}^{{\rm dyn}},J_{9}^{{\rm dyn}}\right]$.
The cost function is composed of two terms, the first is the difference
between the measured and calculated gravity harmonics, and the second
assures that the wind solution does not vary too far from the observed
one at cloud-level. Given the value of $\epsilon_{U}$ and the large
number of coefficients defining the wind latitudinal profile, the
regularization of the wind is very strong, thus ensuring that deviations
from the observed cloud-level wind are allowed only if they result
in a significantly lower value of the cost function. Given an initial
guess for $\overrightarrow{\mathbf{X}_{{\rm C}}}$, a minimal value
of $L$ is searched for using the Matlab function 'fmincon' and taking
advantage of the cost-function gradient that is calculated with the
adjoint of the dynamical model \citet{Galanti2016}. Finally, the
round mean square error (RMSE) we discuss in Table~1 is calculated
by

\begin{equation}
{\rm RMSE}{}_{{\rm gravity}}=\frac{1}{7}\sum_{N}W_{nn}\left(J_{n}^{m}-J_{n}^{o}\right)^{2},\label{eq:RMSE-grav}
\end{equation}
where $n=3,5,6,7,8,9$ and 10, and $N=7$, for Saturn, and $n=3,5,7$
and 9, and $N=4$, for Jupiter.

\section{A combined magnetic-gravity optimization method\label{sec:App-Solve-full}}

The solution for the fully optimized Saturn case is obtained in the
following way. The overall decay function is fixed to the function
obtained in the gravity optimization, and in the optimization process
we look for further modifications in the cloud-level wind so that
in addition to the residual measured gravity field explained by the
thermal model, the residual magnetic field is also explained by the
MFED solution. The MFED model solution is compared to the measured
field by minimizing, in addition to $L_{T}$, the cost function

\begin{equation}
L_{M}=\epsilon_{M}\sum_{n=4}^{11}\frac{1}{\left(e_{n}\right)^{2}}\left(\tilde{g}_{n}^{0}-g_{n}^{0}\right)^{2},\label{eq:cost-magnetic}
\end{equation}
where $\tilde{g}_{n}^{0}$ are the MFED model solutions, $e_{n}$
are the measurements errors (Table~\ref{tab:table1}), and $\epsilon_{M}=10^{5}$
is the weight given the cost function. Note that we take into account
only the Gauss coefficients $g_{4}^{0}$ to $g_{11}^{0}$ that compose
the residual magnetic field. The overall cost function to be optimized
is then

\[
L=L_{T}+L_{M},
\]
and the control vector is now 
\begin{eqnarray}
\mathbf{X}_{{\rm C}} & = & \{\mathbf{X}_{u}\}=\left\{ \left[A_{1}^{{\rm sol}},\cdots,A_{N}^{{\rm sol}}\right]/u_{{\rm nor}}\right\} ,\label{eq:control_vector-augmented-1}
\end{eqnarray}
so that only the parameters composing the cloud-level wind are optimized.
The optimization is done jointly. In each iteration the temporal solution
for the flow structure is used to generate the gravity harmonics with
the thermal wind model and the magnetic coefficients with the MFED
model. Then, the cost function $L$ is calculated and a modified cloud-level
wind is calculated using Matlab 'fmincon'. Finally, the round mean
square error (RMSE) we discuss in Table~2 is calculated by

\begin{equation}
{\rm RMSE}{}_{{\rm magnetic}}=\frac{1}{8}\sum_{n=4}^{11}\frac{1}{\left(e_{n}\right)^{2}}\left(\tilde{g}_{n}^{0}-g_{n}^{0}\right)^{2},\label{eq:RMSE-magnetic-1}
\end{equation}
where $\tilde{g}_{n}^{0}$ is the model solution.

\begin{figure*}[t!]
\begin{centering}
\includegraphics[scale=0.4]{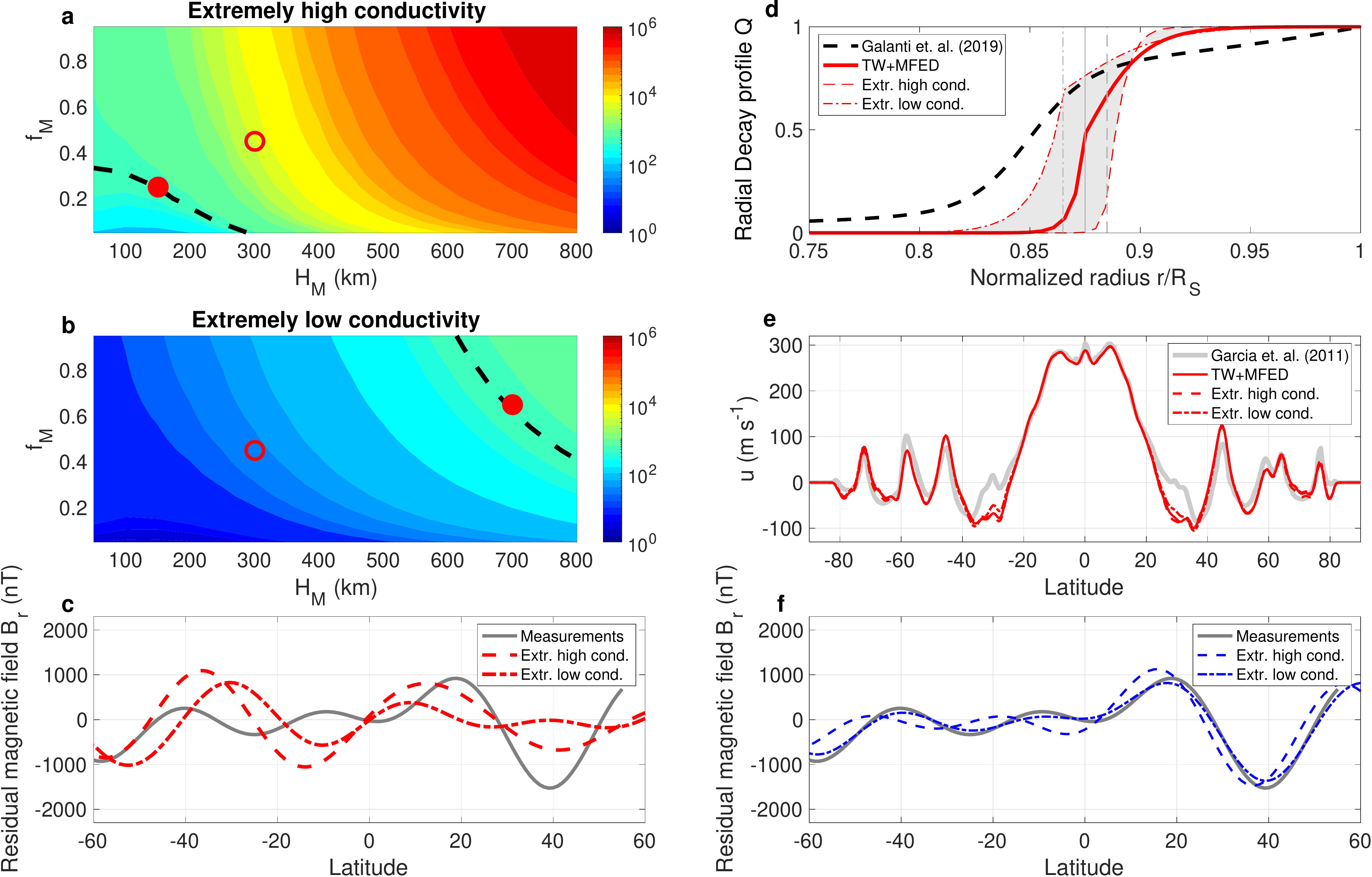}
\par\end{centering}
\centering{}\caption{\textbf{The effect of the uncertainty in the conductivity on the structure
of the flow. }(a) Similar to Figure~\ref{fig:range_Hu_UM}a, but
for a case of extremely high conductivity. The parameter ($H_{M}=150,\,f_{M}=0.25$)
result in a wind induced residual magnetic field fitting the measurements
(full red circle), indicating a very low winds in the conductive region.
(b) Same as (a) but for a case of extremely low conductivity. The
parameters $H_{M}=700,\,f_{M}=0.65$ (full red circle), needed for
a fit to the measurements, represent a stronger and deep winds in
the conductive region. (c) The resulting residual magnetic fields
corresponding to the radial flow profiles chosen (full red circles
in (a) and (b)). (d) the solutions for the full radial profiles of
the extreme cases (red dashed and dashed-dotted), that allow the fitting
the gravity field as well. Also shown are the solution from the main
text (solid red), and the gravity only \citep{Galanti2019a} solution
(black). Note that the transition depth $R_{T}$ is different for
each case, denoting the depth where $\sigma=10^{-2}\,{\rm S\,m^{-1}}.$
(e) The cloud-level wind solutions for the two extreme cases. (f)
The optimal solution of the residual magnetic field for the two extreme
cases (blue dashed and dashed-dotted).\label{fig:range_Hu_UM-extremes-1}}
\end{figure*}

\section{Uncertainties in the MFED model\label{sec:App-Effect-of-uncertainties}}

The electrical conductivity $\sigma$ (see section 2), essential to
the determination of the wind-induced magnetic field, is known within
two order of magnitude in both Jupiter and Saturn \citep{Liu2008,Wicht2019a},
and therefore the effect of its uncertainty on the radial profile
(Figure~\ref{fig:Solutions}) should be evaluated. While the uncertainty
in the electrical conductivity is indeed very large, its effect on
the wind-induced magnetic field is less dramatic, due to the exponential
nature of the its dependence on depth. To illustrate this we examine
two extreme scenarios discussed in the literature \citep{Liu2008}:
one in which the conductivity is an order of magnitude larger and
one in which it is an order of magnitude lower than the mean value.
The results are presented in Figure~\ref{fig:range_Hu_UM-extremes-1}.
In both cases, a radial profile of the flow in the semiconducting
region can be found (Figure~\ref{fig:range_Hu_UM-extremes-1}a,b)
such that the magnitude of the induced residual magnetic field is
similar to the measured one (Figure~\ref{fig:range_Hu_UM-extremes-1}c).
Next, in both cases, a radial profile in the outer region can be found
such that the gravity measurements are also explained (Figure~\ref{fig:range_Hu_UM-extremes-1}d),
with a modified cloud-level wind that is very similar to the solution
with regular conductivity (Figure~\ref{fig:range_Hu_UM-extremes-1}e).
The RMSE for the high and low extreme cases are 0.70 and 0.34, respectively.
Note that in the case with the extreme low conductivity, the gravity
is slightly easier to match than the standard case. Finally, a fully
optimized solution that can explain both the gravity and magnetic
measurements can be found for both extreme cases. The gravity RMSE
are now 0.92 and 0.30 for the high and low cases, respectively. The
magnetic RMSE are now 1.90 and 0.47. Note that, similar to the solutions
without fitting the magnetic field details, it is easier to fit the
magnetic field latitudinal structure with the extreme low conductivity
and somewhat more difficult with the extremely high values. The solutions
for both extreme cases show a similar behavior in the outer region
where the flow is found to be is mostly barotropic. They also show
a similar behavior in having no tail in deep layers. The shift in
the depth of the winds between the two extreme cases (gray region
in Figure~\ref{fig:range_Hu_UM-extremes-1}d) is about 1000~km,
and both solutions are distinctively different from the gravity-only
\citep{Galanti2019a} solution (dashed black). With that, it is evident
that the stronger electrical conductivity results in a sharper decay
of the winds, and the weaker conductivity results in a more moderate
decay that is closer to the shape obtained when fitting the gravity
field only.

Based on the above analysis, we can also discuss even more extremes
values of the conductivity. The solution with the order of magnitude
higher conductivity is already not as good as the regular one (Figure~\ref{fig:range_Hu_UM-extremes-1}f),
and more importantly, further increase in the value will push the
shape of the decay function in the semiconducting region to be unphysical.
This is evident in Figure~\ref{fig:range_Hu_UM-extremes-1}a, where
the values defining the decay function, $H_{M}$ and $f_{M}$, are
already quite close to zero for the extreme case (solid red dot).
Solving the model with another order of magnitude larger conductivity
makes the problem unsolvable. In addition, the high conductivity already
pushes the winds to be completely barotropic in the outer region (Figure~\ref{fig:range_Hu_UM-extremes-1}d,
dashed red line). As we discuss in section~\ref{subsec:A-magnetically-restricted-solution},
this is necessary in order to explain the gravity harmonics, since
this extreme decay function involves less mass (weaker winds) in the
semiconducting zone, and therefore more mass (stronger winds) has
to be included in the outer region. Pushing the decay depth even closer
to the surface, will not allow a fit to the gravity measurements.
As for the extremely weak conductivity, there our model does not pose
any constraint. The weaker the conductivity (or $\alpha$-effect)
is, the more the solution becomes similar to the gravity only solution,
and there can always be a larger value for $H_{M}$ to define the
exponential decay in the semiconducting region.

Another source for uncertainty is the dynamo $\alpha$-effect of which
the latitudinal dependence is not well known \citep{Cao2017a}. However,
modifying the latitudinal dependency of the $\alpha$-effect, as well
as adding the $\gamma$-effect \citep{Kapyla2006}, might add complexity
to the solution but would not change the main results \citep{Cao2017a,Galanti2017e}.
First, the magnitude of $\alpha$ would not be significantly different.
Second, while the latitude dependency of the solution will change,
but as demonstrated here, with minor modification of the cloud-level
wind, the details of the measurements can be explained by the model
solution. Finally, under the assumptions taken in the MFED of small
scale turbulence in the entire semiconducting region, modifying the
overall magnitude of the dynamo $\alpha$-effect is equivalent in
general to changing the value of the conductivity (see equation~\ref{eq:MFED-equations}),
therefore the exploration above should suffice to account for that
uncertainty. With that, it should be noted that a strongly stratified
layer would disconnect the region below it from the winds above, and
render the $\alpha$-effect to be smaller. However, such a stable
layer, if existed, is expected to be in deeper layers \citep{Debras2019}.

\section{Estimates for the magnetic Reynolds number\label{sec:App-Rm}}

\begin{figure}[ht!]
\centering{}\includegraphics[scale=0.35]{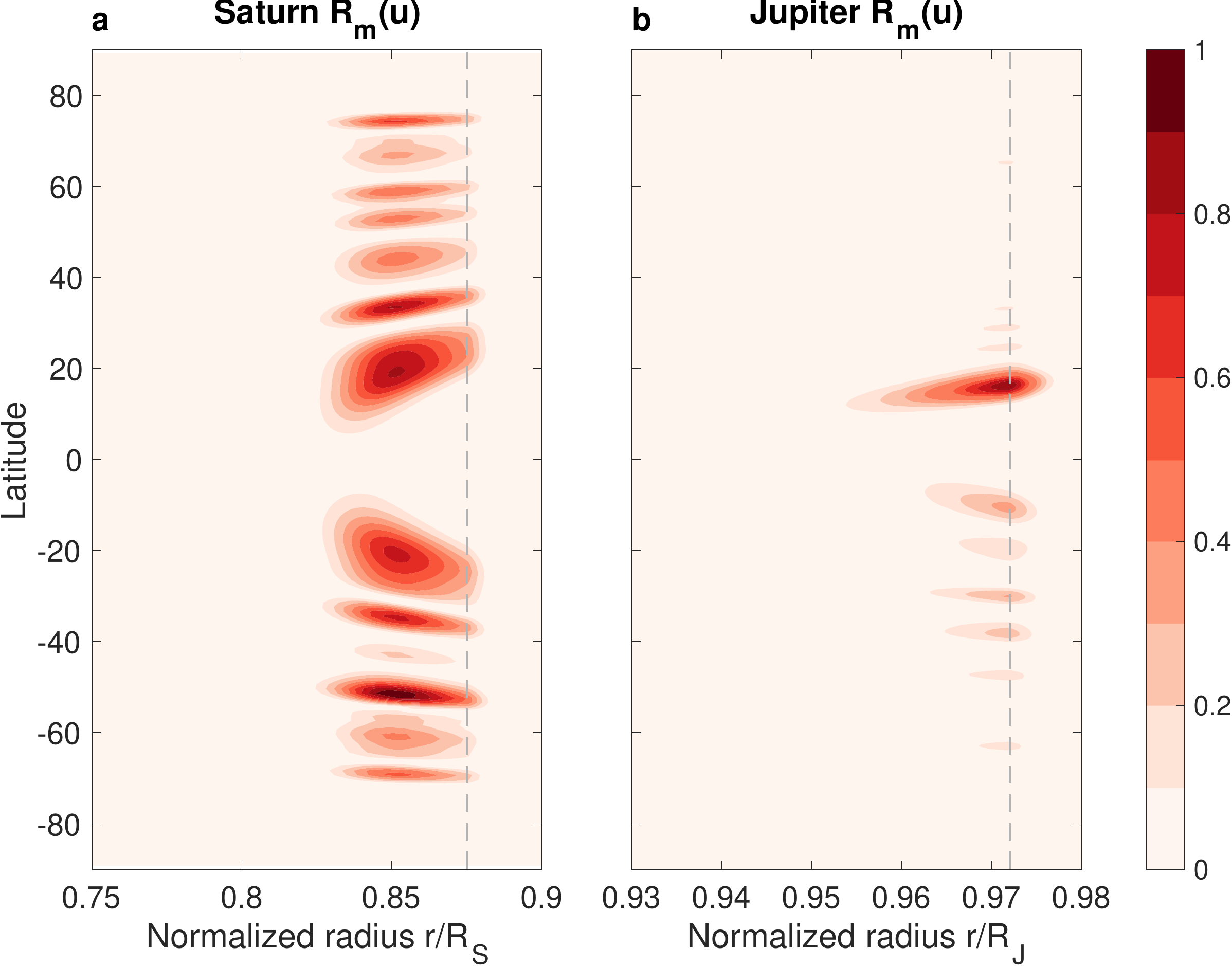}\caption{\textbf{The magnetic Reynolds number $R_{m}(u)$ for the Saturn (a)
and Jupiter (b) solutions, in the semiconducting region.} In the Saturn
case $R_{m}(u)^{max}=0.98$ and in the Jupiter case $R_{m}(u)^{max}=0.88$.
Thus, in both cases the basic requirement that enables using the MFED
balance, $R_{m}(u)<1$, is met.\label{fig:Rm}}
\end{figure}

A basic requirement for using the MFED balance is that the magnetic
Reynolds number would satisfy $R_{m}(u)<1$ \citep{Cao2017a} (it
is defined as $R_{m}(u)=uH_{\sigma}\sigma\mu_{0}$, where $u(r,\theta)$
is the flow velocity, $\sigma$ is the electrical conductivity and
$H_{\sigma}=\sigma/\frac{d\sigma}{dr}$ is the scale height associated
with it). In Figure~\ref{fig:Rm}a we show $R_{m}(u)$ calculated
with the Saturn's optimal flow solution (Figure~\ref{fig:Solutions}a),
where the maximal value in the semiconducting region is 0.98, thus
the condition is satisfied. Note that $R_{m}(u)$ becomes extremely
small close to the lower boundary because the modeled flow goes to
zero there, while in reality the flow strengthens, and so $R_{m}(u)$
might be higher there. This has practically no effect on our results
and conclusion since the gravity harmonics are not sensitive to $O$(1)~m~s$^{-1}$
variations at these depths. Moreover, the wind-induced residual magnetic
field $B_{r}$ can be roughly related to the background field $\mathbf{B_{0}}$
via $B_{r}\sim R_{m}(\alpha)^{{\rm max}}R_{m}(u)^{{\rm max}}\mathbf{|B_{0}|}$,
where $R_{m}(\alpha)=\alpha H_{\sigma}\sigma\mu_{0}$ is the magnetic
Reynolds number associated with the dynamo $\alpha$-effect \citep{Cao2017a}. Since $R_{m}(\alpha)$ reaches a maximal value of 0.25 (at the lower
boundary of the semiconducting region) and $R_{m}(u)<1$ everywhere,
the resulting residual magnetic field is ensured to be much smaller
than the background field (Figure~\ref{fig:range_Hu_UM}b). With
that, the $R_{m}(u)$ associated with our solution is close to 1 in
some regions indicates that the actual flow in the semiconducting
region, especially in its outer part, might be weaker than our solution.

The measured residual magnetic field will then not be solely due to
the winds (see above discussion). In such a case, the decay of the
wind around the 7,000~km depth would be even sharper. A similar analysis
was performed with Jupiter's optimal solution (Figure~\ref{fig:Solutions-Jupiter}a).
The magnetic Reynolds number associated with this solution (Figure~\ref{fig:Rm}b)
has a maximum value of 0.88 in that region, and is less than 0.1 in
most of the domain, thus ensuring that the wind-induced residual magnetic
field will be much smaller than the internal field. Note that other
methods for estimating $R_{m}(u)$ might be used \citep{Wicht2019b},
but this should not affect substantially the results shown here.

\end{document}